\documentclass[apj]{emulateapj}

\def\xmm{{XMM-{\it Newton}}}
\def\chandra{{\it Chandra}}
\def\cp{counterpart}
\def\cps{counterparts}
 
\newcommand{\cgs}{ ${\rm erg~cm}^{-2}~{\rm s}^{-1}$} 
\newcommand{\lum}{\rm erg~s$^{-1}$}
\def\gtrsim{\mathrel{\hbox{\rlap{\hbox{\lower4pt\hbox{$\sim$}}}\hbox{$>$}}}}

\def\lum{{\rm {erg~s$^{-1}$}}}
\usepackage{txfonts}
\usepackage{natbib}
\bibpunct{(}{)}{;}{a}{}{,}
\slugcomment{Accepted for publication in ApJS on May 21, 2012}

\shorttitle{\chandra\ COSMOS: optical properties}
\shortauthors{F. Civano et al.}

\begin{document}

 \title{The \chandra\ COSMOS Survey: III. 
   Optical and Infrared Identification of X-ray Point Sources}

\author{ F. Civano\altaffilmark{1}, M. Elvis\altaffilmark{1}, M. Brusa\altaffilmark{2}, 
A. Comastri\altaffilmark{3}, M. Salvato\altaffilmark{2,4}, G. Zamorani\altaffilmark{3}, 
T. Aldcroft\altaffilmark{1}, A. Bongiorno\altaffilmark{2,7}, P. Capak\altaffilmark{5}, N. Cappelluti\altaffilmark{3}, M. Cisternas\altaffilmark{6}, 
F. Fiore\altaffilmark{7}, A. Fruscione\altaffilmark{1}, H. Hao\altaffilmark{1}, J. Kartaltepe\altaffilmark{8}, A. Koekemoer\altaffilmark{9}, 
R. Gilli\altaffilmark{3}, C. D. Impey\altaffilmark{10},
G. Lanzuisi\altaffilmark{1,7}, E. Lusso\altaffilmark{3}, V. Mainieri\altaffilmark{11}, T. Miyaji\altaffilmark{12}, 
S. Lilly\altaffilmark{13}, D. Masters\altaffilmark{14}, S. Puccetti\altaffilmark{15}, K. Schawinski\altaffilmark{16}, N. Z. Scoville\altaffilmark{5}, 
J. Silverman\altaffilmark{17}, J. Trump\altaffilmark{10}, M. Urry\altaffilmark{16}, C. Vignali\altaffilmark{3,18}, N. J. Wright\altaffilmark{1}
}
 
\altaffiltext{1}{Harvard Smithsonian Center for astrophysics, 60 Garden St., Cambridge, MA 02138, USA}
\altaffiltext{2}{Max Planck Institut f\"ur extraterrestrische Physik Giessenbach strasse 1, D--85748 Garching, Germany}
\altaffiltext{3}{INAF-Osservatorio Astronomico di Bologna, via Ranzani 1, I-40127 Bologna, Italy}
\altaffiltext{4}{Excellence Cluster, Boltzmannstrass 2, Garching D--85748, Germany} 
\altaffiltext{5}{California Institute of Technology, 1201 East California blvd, Pasadena, CA 91125, USA}
\altaffiltext{6}{Max-Planck-Institut f\"ur Astronomie, Konigstuhl 17, D-69117 Heidelberg, Germany} 
\altaffiltext{7}{INAF-Osservatorio Astronomico di Roma, via Frascati 33, Monteporzio-Catone (Roma), I-00040, Italy}
\altaffiltext{8}{Institute for Astronomy, 2680 Woodlawn Dr., University of Hawaii, Honolulu, HI 96822, USA}
\altaffiltext{9}{Space Telescope Science Institute, 3700 San Martin Drive, Baltimore, MD 21218, USA}                                     
\altaffiltext{9}{Department of Physics and Astronomy, University of California, 900 University Ave, Riverside, CA 92521, USA}
\altaffiltext{10}{Steward Observatory, University of Arizona, 933 North Cherry Avenue, Tucson, AZ 85721, USA}
\altaffiltext{11}{ESO, Karl-Schwarzschild-Strasse 2, 85748 Garching, Germany} 
\altaffiltext{12}{Universidad Nacional Autonoma de Mexico-Ensenada, Km. 103 Carretera Tijuana-Ensenada, 22860 Ensenada, Mexico}
\altaffiltext{13}{Institute of Astronomy, Swiss Federal Institute of Technology, 8093, Zurich, Switzerland}
\altaffiltext{14}{Observatories of the Carnegie Institution of Washington, Pasadena, CA 91101, USA}
\altaffiltext{15}{ASI Science Data Center, via Galileo Galilei, 00044 Frascati, Italy} 
\altaffiltext{16}{Department of Physics, Yale University, New Haven, CT 06511, USA}
\altaffiltext{17}{Institute for the Physics and Mathematics of the Universe (IPMU), University of Tokyo, Kashiwanoha 5-1-5, Kashiwa, Chiba 277-8568, Japan} 
\altaffiltext{18}{Dipartimento di Astronomia, Universit\'a di Bologna, via Ranzani 1, 40127, Bologna, Italy}


\begin{abstract}
The {\em Chandra} COSMOS Survey (C-COSMOS) is a large, 1.8~Ms, \chandra\ 
program that has imaged the central 0.9 deg$^2$ of the COSMOS
field down to limiting depths of 1.9$\times$10$^{-16}$~erg~cm$^{-2}$~s$^{-1}$ in the soft (0.5--2~keV)
band, 7.3$\times$10$^{-16}$~erg~cm$^{-2}$~s$^{-1}$ in the hard
(2--10~keV) band, and 5.7$\times$10$^{-16}$~erg~cm$^{-2}$~s$^{-1}$ in
the full (0.5--10~keV) band.
In this paper we report the i, K and 3.6$\mu m$ identifications 
of the 1761 X-ray point sources. 
We use the likelihood ratio technique to derive the association of 
optical/infrared \cps\ for 97\% of the X-ray sources. 
For most of the remaining 3\% , the presence of multiple \cps\ or the faintness of 
the possible \cp\ prevented a unique association. For only 10 X-ray sources 
we were not able to associate a \cp, mostly due to the presence of a very bright 
field source close by. Only 2 sources are truly empty fields.
The full catalog, including spectroscopic and photometric redshifts and classification 
described here in detail, is available on-line.
Making use of the large number of X-ray sources, we update the 
``classic locus'' of AGN defined 20 years ago in soft X-ray surveys and 
define a new locus containing 90\% of the AGN in the survey with full band luminosity $>$10$^{42}$\lum. 
We present the linear fit between the total {\it i} band magnitude and the X-ray flux in the soft and hard band, 
drawn over 2 orders of magnitude in X-ray flux, obtained using the combined C-COSMOS and XMM-COSMOS samples. 
We focus on the X-ray to optical flux ratio (X/O) and we test its known correlation with redshift and luminosity, 
and a recently introduced anti-correlation with the concentration index (C). We find a strong anti-correlation (though the dispersion is of the order of 0.5 dex) between C and X/O, 
computed in the hard band, and that 90\% of the obscured AGN in the sample with morphological information live 
in galaxies with regular morphology (bulgy and disky/spiral), suggesting that secular processes govern a significant fraction of the BH growth 
at X-ray luminosities of 10$^{43}- 10^{44.5}$\lum. 
We also investigate the degree of obscuration of the sample, using the 
hardness ratio and we compare the X-ray color with the near-infrared-to-optical one.

\end{abstract}

\keywords{galaxies: active -- surveys -- X-rays:galaxies}

\section{Introduction}
 
There is now much evidence showing that galaxies and their central supermassive black holes (SMBHs) 
undergo closely coupled evolution. The masses of SMBHs in the nuclei of most
nearby bulges are tightly proportional to bulge luminosity and velocity
dispersion (Magorrian et al. 1998, Gebhardt et al. 2000, Ferrarese \& Merrit 2000, Merloni et al. 2010). 
Most SMBH growth occurs during their active  phases (the ``Soltan
argument'', Soltan 1982), implying that most bulges went through an  Active Galactic Nuclei (AGN) phase. Galaxies and AGN 
exhibit co-ordinated cosmic ``downsizing'': the star-formation of massive galaxies peaks at z$\sim$2 (Cimatti et al. 2006), and BH growth, 
as traced by quasar luminosity function (Hasinger et al. 2005, Silverman et al. 2008), peaks in the same redshift range (z=2-3), 
while the star-formation of lower mass galaxies peaks at z=1-1.5, as do lower-mass and lower-luminosity AGN (La Franca et al. 2005, Bongiorno et al. 2007). 
 SMBH growth seems to be connected with galaxy growth, but the details of this mutual relation remain still poorly understood (i.e., Merloni \& Heinz 2008).

To fully understand how co-evolution works requires the measurement of at least 7
parameters in large samples of AGN: (1) SMBH mass (M$_{BH}$, using broad emission line widths and optical luminosity), (2) SMBH growth rate 
(from bolometric luminosity)
(3) galaxy mass (M$_{star}$, via optical to infrared spectral energy distributions, SEDs, fitting), (4) star formation rate (SFR, using far infrared or UV emission,
narrow emission lines in optical spectra), (5) AGN host galaxy morphology (high resolution imaging), (6) galaxy environment
(using spectroscopic or reliable photometric redshifts) and (7) Dark Matter halo mass (lensing studies). To measure all the above quantities, 
a complete sample with spectroscopic and radio to X-ray photometric data is necessary. To assemble this kind of sample is not an easy task.

The selection of AGN at X-ray wavelengths does not suffer from the 
heavy contamination by non-active (mainly star-forming) galaxies that affects infrared or optical surveys (Donley et al. 2008, 2012). Moreover, 
X-ray surveys are efficient in 
selecting low-luminosity and obscured AGN up to equivalent hydrogen column densities, N$_H$, of 10$^{24}$ cm$^{-2}$ (Compton thick regime). Therefore, X-ray selected samples of AGN, 
coupled with multiwavelength data, are ideal to cleanly derive the bolometric output and the accretion rate of the BHs in the majority of AGN.
For these reasons, over the last 10 years, deep and wide X-ray surveys have been carried out with the \chandra\ and \xmm\ satellites to 
generate samples of X-ray emitting AGN, coupled with deep multiwavelength spectroscopic and photometric coverage to provide all
the above parameters for large samples over large volumes.

\chandra\ and \xmm\ contiguous extragalactic surveys are, like a wedding cake, layered in increasing depth but 
decreasing area (see Figure 5 of Elvis et al. 2009), from the wide/shallow XBootes survey (9 deg$^2$, Murray et al. 2005) to the ultra-deep/narrow survey of the 4 Ms 
\chandra\ Deep Field South (CDFS, 0.1 deg$^2$, Xue et a. 2011). 
The major advantage of ultra-deep surveys is that they are able to detect the X-ray emission of non-active galaxies at 
medium-high redshifts (Luo et al. 2011, Lehmer et al. 2012), although they do not probe a large cosmological volume and thus 
contain relatively small numbers of sources at any given redshift. 
Larger areas are required to uniformly sample the parameter space 
(luminosity and redshift), obtain large samples of rare sources (e.g., recoiling SMBH, Civano et al. 2010)
and measure large-scale structures in the universe (Hickox et al. 2011, Allevato et al. 2011, Capak et al. 2011) 
and hence determine the relationship between galaxy evolution and local environmental density.
 For a complete picture, it is clearly necessary to explore the parameter space between the ultra-deep and ultra-wide surveys. Eventually, 
 the best approach to trace the full population of AGN at all redshifts and all luminosities is to merge the data coming from all the 
 layers of the wedding cake (see e.g., Fiore et al. 2012). 

The \chandra-COSMOS project (C-COSMOS; Elvis et al. 2009, Paper I hereinafter) hits a sweet spot among the wedding cake layers (Fig.5 in
Elvis et al. 2009): deep enough (1.9$\times$10$^{-16}$~erg~cm$^{-2}$~s$^{-1}$ in the 0.5--2~keV band) to find obscured AGN with 
optical galaxy continua and wide enough (0.9 deg$^2$) to have large samples (1761 X-ray point-like sources) and find unusual, 
rare objects (Civano et al. 2010, Civano et al. 2012, Capak et al. 2011). Yet, the C-COSMOS sources are bright enough that virtually all X-ray sources 
can be identified and followed up across their SEDs, especially with optical or near-IR spectroscopy, using the vast Cosmic Evolutionary Survey 
(COSMOS, Scoville et al. 2007) multiwavelength photometric and spectroscopic database 
(Schinnerer et al. 2007, Sanders et al. 2007, Taniguchi et al. 2007, Capak et al. 2007, Koekemoer et al. 2007, Zamojski et al. 2007, Lilly et al. 2007, 
Trump et al. 2007). 
Largely thanks to the low background of \chandra, the C-COSMOS survey reaches 4 times fainter fluxes 
than XMM-COSMOS (Hasinger et al. 2007, Cappelluti et al. 2009, Brusa et al. 2010) in both the 0.5-2 keV and 2-8 keV bands. The heavily overlapping 
``dense tiling'' observation strategy, now also used in other \chandra\ surveys (e.g., Drake et al. 2009), combined with the sub-arcsecond on-axis \chandra\ point 
spread function (PSF, Weisskopf et al. 2002), provides a rather uniform 
sensitivity in C-COSMOS (see Paper I and Puccetti et al. 2009, Paper II, for details). Thus, the high-resolution area is maximized to resolve sources 
with a sub-arcsecond position accuracy, a key ingredient for the best identification of the optical counterparts of the X-ray sources, to find AGN pairs 
(Silverman et al. 2011, Cunningham et al. in prep.) and off-nuclear sources (Mainieri et al. 2010). 

Historically, various classes of X-ray emitters have been characterized by different
values of the X-ray-to-optical flux ratio (hereinafter X/O = $log(f_X/f_{opt})$) which provides a first 
indication of the source classification (Maccacaro et al. 1988). 
Originally, the X/O ratio was defined in the soft band of the Einstein Observatory (0.3-3.5 keV) at fluxes of 10$^{-13}$ to 
10$^{-11}$ \cgs\ and later modified to the lower energy ROSAT band (0.1-2.4 keV); 
the majority of spectroscopically identified AGN in soft X-ray surveys were characterized by X/O=0$\pm$1.

With the advent of \chandra\ and \xmm, the same relation has been used in harder bands (Hornschemeier et al. 2001, 
Alexander et al. 2001, Fiore et al. 2003, Della Ceca et al. 2004, Cocchia et al. 2007). The optical identification of the sources found in deep and medium
surveys (\chandra\ Deep Fields, Hellas2XMM, XMM-COSMOS) confirms the trend observed at bright fluxes 
but also shows a large scatter around the median value in both the hard and soft bands (Brandt \& Hasinger 2005). 
Obscured AGN (N$_H>$10$^{22}$ cm$^{-2}$) tend to populate the upper part of this plot and 
have $X/O>1$ (Perola et al. 2004, Civano et al. 2005, Brusa et al. 2010), while normal galaxies, detected at very faint 
X-ray fluxes, have $X/O<-2$ (Xue et al. 2011). In addition to obscured AGN, unobscured
X-ray Bright Optically Normal Galaxies (XBONGs) also have $X/O>1$ (e.g. Comastri et al. 2002, Civano et al. 2007, Trump et al. 2009c). 
The lack of information for faint X-ray and optical sources, plus their high X/O ratio, 
led the scientific community to name this kind of sources ``extreme'' or ``unconventional'' (Comastri et al. 2003, Mignoli et al. 2004), 
in particular when the X/O is defined in the hard X-ray band.

In this paper (Paper III), we present the catalog of optical and infrared \cps\ of 
\chandra-COSMOS sources, presented in Paper I, and we analyze some their observed properties. The paper is organized as follows: 
Section 2 describes the datasets used in this work; Section 3 and 4 explain the details of the method used for the identifications and 
the results obtained; in Section 5 the positional and magnitude distributions are presented; 
in Section 6 we show the spectroscopic and photometric redshifts along with the spectral and SED classification; in Section 7 we present the identification catalog; 
in Section 8 the optical and infrared properties are compared with the X-ray ones; in Section 9 we summarize the results of this work. 

We assume a cosmology with H$_0$ = 70~km~s$^{-1}$~Mpc$^{-1}$, 
$\Omega_M$ = 0.3 and $\Omega_{\Lambda}$= 0.7. The AB magnitude system is used in this paper if not otherwise stated.

\section{Identification Datasets}

The C-COSMOS X-ray source catalog used in this work comprises 1761 X-ray point sources 
detected down to a maximum likelihood threshold detml$=$10.8 in at least one band (0.5-2, 2-8 or 0.5-8 keV). 
This likelihood threshold corresponds to a probability of $\sim 5 \times 10^{-5}$ that a catalog source is instead a 
background fluctuation (Paper II). At this threshold, the flux limit reached in the survey is 
5.7$\times$10$^{-16}$~erg~cm$^{-2}$~s$^{-1}$ in
the full band (0.5--10~keV), 1.9$\times$10$^{-16}$~erg~cm$^{-2}$~s$^{-1}$ in the
soft band (0.5--2~keV) and 7.3$\times$10$^{-16}$~erg~cm$^{-2}$~s$^{-1}$ in the
hard band (2--10~keV, extrapolated from the observed 2-7 keV band). Of the 1761 sources, 922 are detected in all three bands (full, soft, hard), 474
are detected in the full and the soft band, 
257 are detected in the full and the hard band, 73 only in the full, 32 only in
the soft and 3 only in
the hard band. Upper limits (3 $\sigma$ values) in each band have been computed for the sources detected in only 
one or two bands (see Paper II). 

We identify counterparts of the C-COSMOS sources in the {\it i}-band (0.76$\mu m$) using  
the Subaru photometric catalog (Capak et al. 2007, hereinafter ``optical catalog''), in the K-band (2.15$\mu m$) using 
the CFHT/Megacam catalog (McCracken et al. 2010) and at 3.6$\mu m$ using the Spitzer IRAC catalog (Sanders et al. 2007).
The full optical catalog contains about 1.5 million sources down to a magnitude limit {\it i}$\sim$ 27 (AB) at 
a signal-to-noise ratio (S/N$>$5). The K-band catalog contains $\sim$500,000 sources detected at S/N$>$5 to 
a magnitude limit of 23.5 (AB).
The IRAC catalog contains $\sim$350,000 sources at 3.6~$\mu m$ to 23.9 (AB, corresponding to $\sim$1$\mu$Jy) at a S/N $>$5. 
 
In the second part of the identification process, we also made use of the ACS/HST images of the COSMOS field 
(Scoville et al. 2007, Koekemoer et al. 2007) to visually check our identifications, taking advantage of the ACS PSF, 
of the accuracy of the positions and of the depth of the 
observations (I$_{F814W}\sim$27.8 AB mag, 5 $\sigma$ for an optimally extracted point source). 
A new, revised photometric catalog, which includes the photometry in all the 25 optical/NIR broad-, 
intermediate- and narrow-bands filters, with improved source detection\footnote{The catalog is publicly available at 
\url{http://irsa.ipac.caltech.edu/data/COSMOS/tables/photometry/} and supercedes the catalog reported by Capak et al. (2007). 
The photometry is computed at the position of the  {\it i}$^*$-band image, using Sextractor (Bertin \& Arnouts 1996) 
in dual mode. The same catalog is used in Ilbert et al. (2009, 2010) and Salvato et al. (2009, 2011).} has been used to match the final identifications.

\section{X-ray Source Identification Method}

\subsection{Method}
We used the likelihood ratio technique of Sutherland \& Saunders (1992), following the procedure described by 
Brusa et al. (2005) which, when applied to the identification of the 
XMM-COSMOS sample (Brusa et al. 2007, 2010) yielded a percentage of ``reliable identifications'' greater than 80\%. 
 With respect to simpler matching approaches, based only on the positional offset between sources in different catalogs, 
the likelihood ratio technique has the advantage of taking into account, for each possible counterpart, the probability that it is 
a real or a spurious identification, allowing a much better statistical characterization of the final sample of identifications.
This powerful statistical method has become common recently and has been applied first to radio surveys (Ciliegi et al. 2005) and 
to several medium and faint \chandra\ and \xmm\ surveys 
(Cardamone et al. 2008, Laird et al. 2009, Aird et al. 2010, Luo et al. 2010, Brusa et al. 2010, Xue et al. 2011, Rovilos et al. 2011). \\

Identifying the true \cp\ of \chandra\ sources is easier than for \xmm\ because of the narrower \chandra\ PSF, 
therefore excellent positional accuracy (on average $<$0.5$^{\prime\prime}$, Paper II), 
and the tighter \chandra\ encircled energy 
fraction\footnote{The encircled energy fraction is the two-dimensional integral of the PSF.} (typically 90\% at 0.5$^{\prime\prime}$ 
on-axis to 5$^{\prime\prime}$ at 12$^{\prime}$ off-axis; Weisskopf et al. 2002). 
X-ray sources at C-COSMOS fluxes have, on average, redder colors than brighter X-ray sources (Brandt \&\ Hasinger 2005)
and thus the likelihood ratio in 
near-infrared bands should produce a high rate of identifications. In this paper we therefore give equal weight 
to the redder bands (K, 3.6$\mu m$) as to the optical ({\it i}), so that we can maximize the number of identified sources,
and also compare the identification efficiency at different wavelengths. This approach is possible only because 
the depth of the K and IRAC 3.6$\mu m$ COSMOS observations nicely matches the X-ray depth for a typical AGN spectral energy distribution 
(SED, e.g. Elvis et al. 1994, Elvis et al. 2012 sub.). Deep X-ray surveys (e.g., 2 or more Megaseconds in the CDFS) reveal, besides 
a population of bright normal galaxies, 
a population of fainter X-ray sources which will be hard to follow up across the SED with the currently available optical or near-IR capabilities.  
  
We can summarize the four steps of our identification procedure as follows: 
\begin{enumerate}
\item run the likelihood ratio technique separately in the 3 different bands ({\it i}, K and 3.6~$\mu m$; Section 3.2);
\item find appropriate parameters (threshold, normalization) in order to identify the sources and define the level of confidence of the association (Section 3.3);
\item combine the results obtained in the 3 bands (Section 4);
\item confirm the uncertain cases visually checking the HST ACS image at the position of the X-ray source (Section 4.1 and 4.3).
\end{enumerate}

\subsection{Likelihood Ratio Definition}

For a given optical or infrared candidate with magnitude {\it m} and 
positional offset {\it r} from the X-ray source position, the likelihood 
ratio $LR$ is defined as the ratio between the probability
that the source is the correct identification and the corresponding 
probability for a background, unrelated object:
\begin{equation}
LR=\frac{q(m) f(r)}{n(m)},
\end{equation}
where {\it q(m)} is the expected magnitude {\it m} distribution function (normalized to one) of the real optical \cp\ 
candidates, 
{\it f(r)} is the probability distribution 
function of the positional errors (assumed to be a two-dimensional Gaussian) and {\it n(m)} is the surface density 
of background objects with magnitude {\it m}.

The distribution of the local background objects, {\it n(m)}, was computed 
from each of the three input catalogs using the objects 
within a 5$^{\prime\prime}$--30$^{\prime\prime}$ annulus around each X-ray source. 
We chose a 5$^{\prime\prime}$ inner radius in order to avoid the presence of true \cps\ in the background 
distribution, and a 30$^{\prime\prime}$ outer radius to exclude the \cps\ of other nearby 
X-ray sources. 
In the case of X-ray source pairs (178 pairs with relative distances $<$15$^{\prime\prime}$ and 400 with distances $<$30$^{\prime\prime}$), the 
outer radius could contain the \cp\ of a nearby X-ray source. However, 
the number of background sources in the annuli is sufficiently large ($\sim$100 sources on average in the K band down to 
22 mag and $\sim$300 sources in the {\it i} band down to 25 mag) to make the presence of a nearby object's \cp\ unimportant.

\begin{figure}
\centering
\includegraphics[width=0.49\textwidth]{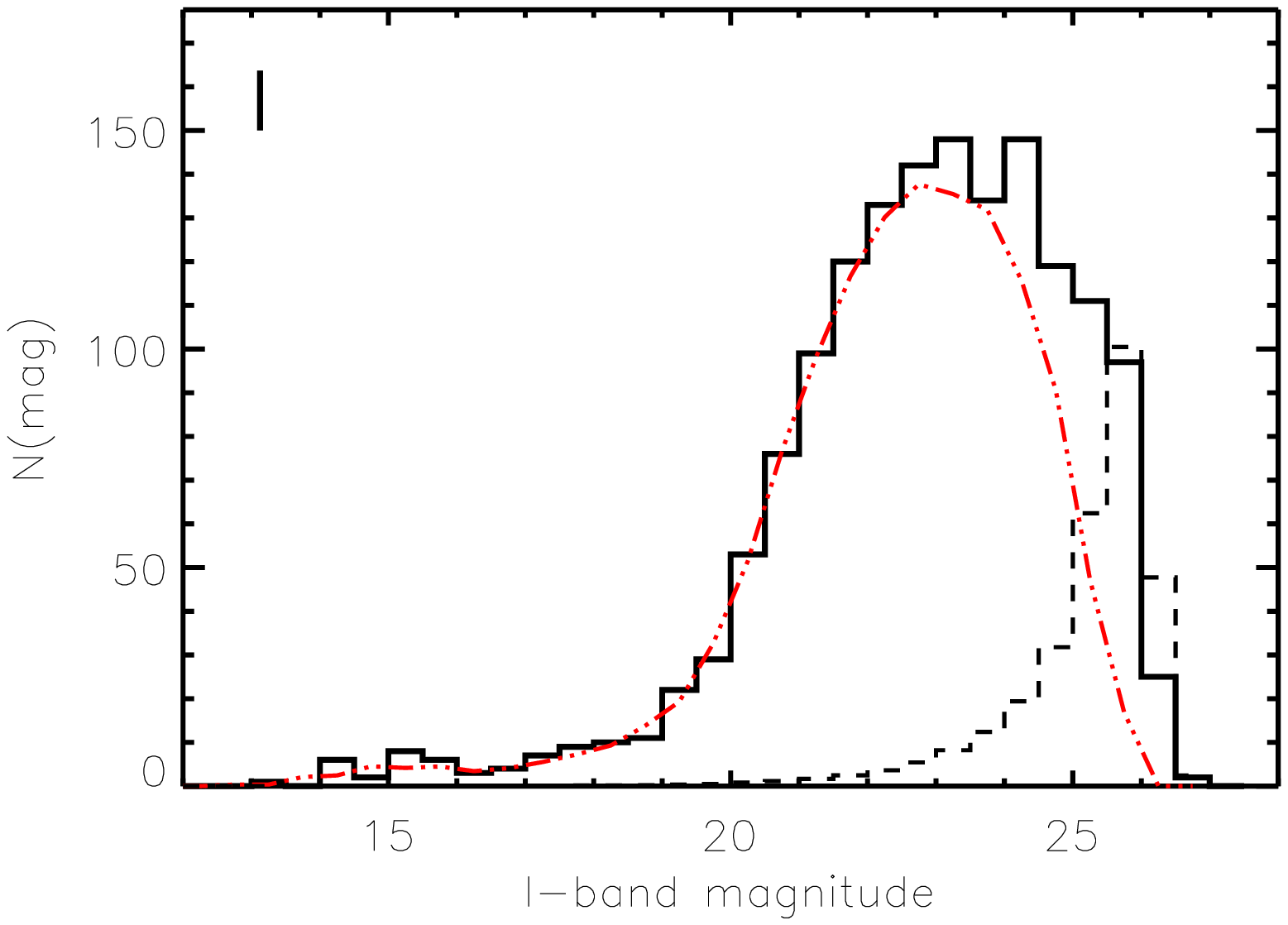}
\includegraphics[width=0.49\textwidth]{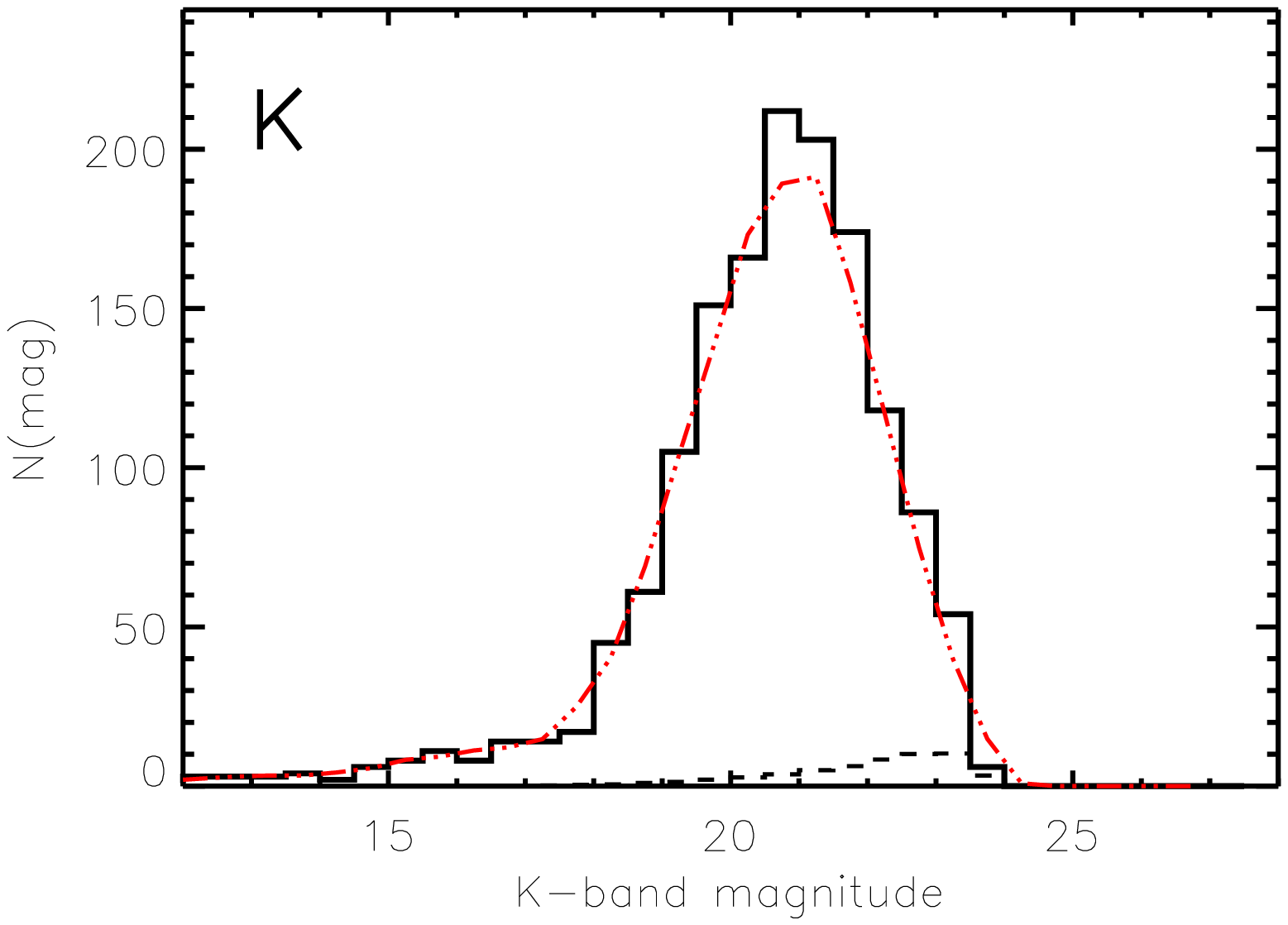}
\includegraphics[width=0.49\textwidth]{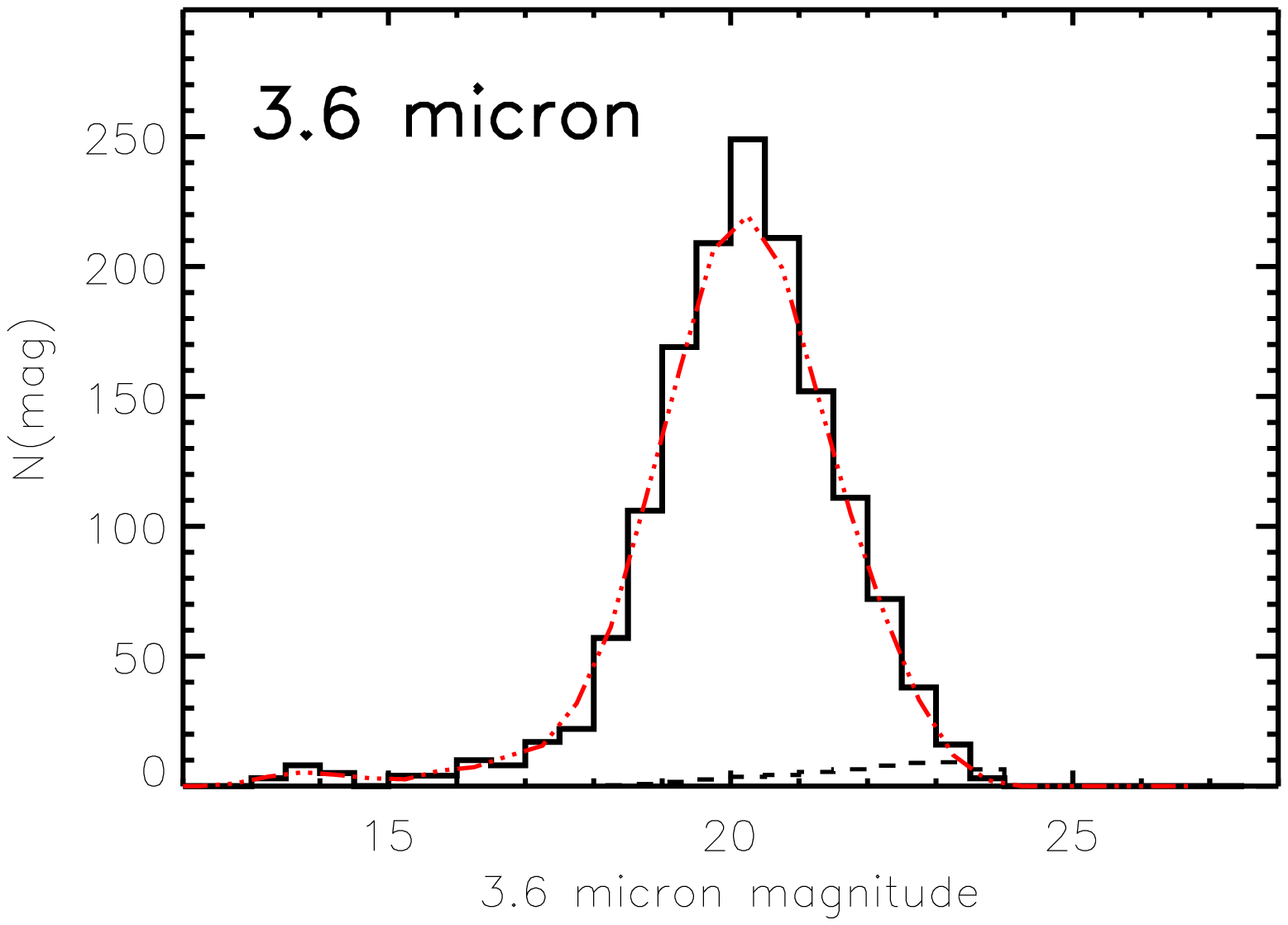}
\caption{\small Observed AB magnitude distribution of all the {\it i}-band, K-band and 3.6$\mu m$ band (from top to bottom) objects 
within a radius of 1$^{\prime\prime}$ around the 1761 X-ray sources (solid black histogram) together with the expected 
distribution of background objects ({\it n(m)}, black dashed histogram). The difference between the two distributions 
(red dash-dotted line) corresponds to the expected magnitude distribution of \cps\ ({\it q'(m)}, smoothed here for plotting purposes).}
\label{expected}
\end{figure}

The function {\it q(m)} has been estimated from our data as follows. In this work,
we neglect the possible X-ray flux dependence of {\it q(m)} and assumed a universal
optical/infrared magnitude distribution for all X-ray sources.
 We first computed {\it q'(m)}= [number of sources with magnitude m within 1$^{\prime\prime}$]$-$ [expected number of background sources with 
magnitude m in a 1$^{\prime\prime}$ circle].
The choice of a 1$^{\prime\prime}$ radius is dictated by the requirement of maximizing the statistical significance of the over-density 
around the X-ray sources. A smaller radius would include in the analysis only a fraction of the true identifications and the q(m) distribution 
would be more affected by Poissonian noise. A larger radius would increase the number of background sources.

In Figure \ref{expected}, the observed magnitude distribution of the objects
in the 3 catalogs ({\it i}, K and 3.6 $\mu$m)  within a radius of 
1$^{\prime\prime}$ around each X-ray source (solid histogram) is plotted together with the expected distributions 
of background objects in the same area (dashed histogram). The difference between these two distributions is the expected 
distribution of the \cps\ ({\it q'(m)}, red curve) before normalization.  
The distribution of background sources in the {\it i} band (Fig. \ref{expected}, top panel) is more prominent than in 
the K or 3.6$\mu m$, because of the fainter magnitude limit of the optical catalog (the number counts of galaxies in {\it i} band 
strongly increase at faint magnitude, see e.g. Ilbert et al. 2010). \\

Then, $q(m)=$const$\times q^\prime(m)$ is normalized such that
$\int_{-\infty}^{+\infty}q(m)dm=1$. In practice, since we have a magnitude limit $m_{\rm lim}$,
we normalized in such a way that 
$Q=\int_m q(m) dm=$ (number of X-ray sources identified in each band / the total number of sources of the X-ray sample). 
More details are given in $\S$3.3.\\

For the probability distribution of positional errors, {\it f(r)}, we adopted a 
Gaussian distribution with 
standard deviation, $\sigma$=$\sqrt{\sigma_{X}^2+ \sigma_{opt}^2}$, where $\sigma_{opt}$ is the positional uncertainty in the 
3 optical and infrared bands. $\sigma_{X}$ is taken from column 5 in Table 7 of Paper I (see also 
Fig. 12 in Paper II).
The adopted $\sigma_{opt}$ is 0.2$^{\prime\prime}$ for the K band (McCracken et al. 2010), 0.3$^{\prime\prime}$ in {\it i} 
band (Capak et al. 2007) and 0.5$^{\prime\prime}$ at 3.6$\mu m$ (Sanders et al. 2007).\\

Having determined the values of {\it q(m)}, {\it f(r)} and {\it n(m)} for each band, using Eq.~1 we 
computed the $LR$ value for all the sources within 5$^{\prime\prime}$ of the 1761 
X-ray centroids in each of the 3 catalogs (a total of 10972 sources for the I-band, 2992 sources for the K-band and 2604 for the IRAC 3.6~$\mu m$ band).

\subsection{Threshold Choices}
 
Once the $LR$ values have been computed for all the candidates, one has to choose the best likelihood threshold 
value ($L_{\rm th}$) for $LR$ to discriminate between spurious and real 
identifications. The choice of $L_{\rm th}$ depends on two factors: first, $L_{\rm th}$ 
should be small enough to avoid missing many real 
identifications, so that the completeness of the sample is high; 
secondly, $L_{\rm th}$ should be large enough to keep the number 
of spurious identifications low, in order 
to increase the reliability of the identification.  

The reliability takes into account the possibility of having multiple optical candidates for the same X-ray source.
The reliability $R_j$ for object {\it j} being the correct identification is:
\begin{equation}
R_j=\frac{(LR)_j}{\Sigma_i (LR)_i + (1-Q)}
\end{equation}
where the sum is over the set of all candidates for source {\it j}. 
The reliability parameter ({\it R}) for the total sample is defined as the ratio 
between the sum of the reliability of all the sources identified as possible \cps\ and the total 
number of sources with $LR >L_{\rm th}$ (R= $N_{ID}/N_{LR>L_{\rm th}}$). 

The completeness parameter ({\it C}) of the total sample is 
defined as the ratio between the sum of the reliability of all the sources identified as possible \cps\  
and the total number of X-ray sources (C= $N_{ID}/N_X$).

Following Brusa et al. (2007), we defined $L_{\rm th}$ as the likelihood ratio which maximizes the quantity (R+C)/2. 
We applied this criterion to the three input catalogs ({\it i}, K and 3.6$\mu m$) and the resulting 
$L_{\rm th}$ are 0.4, 0.9, 0.47, respectively. 

\begin{figure}[t]
\centering
\includegraphics[width=0.5\textwidth]{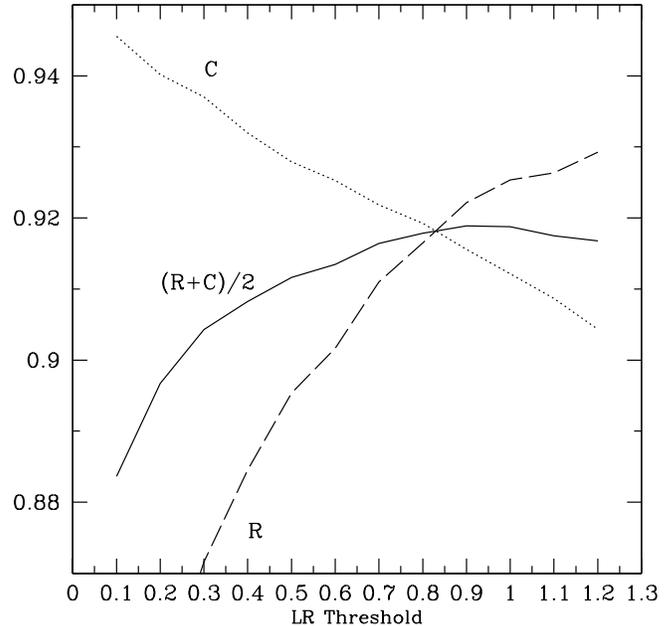}
\caption{\small Completeness (dotted line) and reliability (dashed line) parameter and (C+R)/2 (solid line) of our sample in 
the K band with a normalization value {\it N}=0.9. The $L_{\rm th}$ chosen corresponds to the value where the 
(C+R)/2 (solid line) quantity is maximized but still close to the intersection of the 3 curves. }
\label{CN}
\end{figure}

The corresponding sample completeness and reliability are 
C=0.85 and R=0.88 for {\it i}, C=0.90 and R=0.92 for $K$, and C=0.96 and R=0.96 for 3.6$\mu m$.
As an example, in Figure \ref{CN}, {\it C}, {\it R} and {\it (R+C)/2} versus $L_{\rm th}$ are reported for the 
K-band catalog, using 0.9 as normalization for {\it q'(m)} to convert into {\it q(m)}. {\it (R+C)/2} is quite flat over 
0.7$< L_{\rm th} <$ 1.2. The value chosen corresponds to the point where {\it (C+R)/2} is maximized and it is close to where 
{\it C}, {\it R} and {\it (C+R)/2} curves intersect.

 It is possible to make a direct comparison with the likelihood threshold values used by Brusa et al. (2007) for XMM-COSMOS 
because the same input optical catalog was used. Thanks to the better \chandra\ positional accuracy, 
at the same LR threshold in {\it i} band ($L_{\rm th}$=0.4) a higher completeness and reliability can be 
achieved in \chandra\ than in XMM (C=0.77 and R=0.69 in Brusa et al. 2007, versus C=0.85 and R=0.88 here).  
A lower LR threshold in {\it i} band was used for the final XMM-COSMOS catalog (Brusa et al. 2010, $L_{\rm th}$=0.26), however, 
its completeness and reliability cannot be compared with the C-COSMOS one, as the \chandra\ COSMOS associations 
have been used to resolve some {\it ambiguous} identifications in the XMM-COSMOS catalog.\\

\section{X-ray Source Identification Results}
In this section we present the various steps that we followed to define the best associations after running the likelihood ratio analysis. 
We first used the output of the likelihood ratio, then we merged the results in the 3 bands, we inspected the associations 
and listed the final \cps. 

\begin{figure*}
\centering
\includegraphics[width=0.9\textwidth]{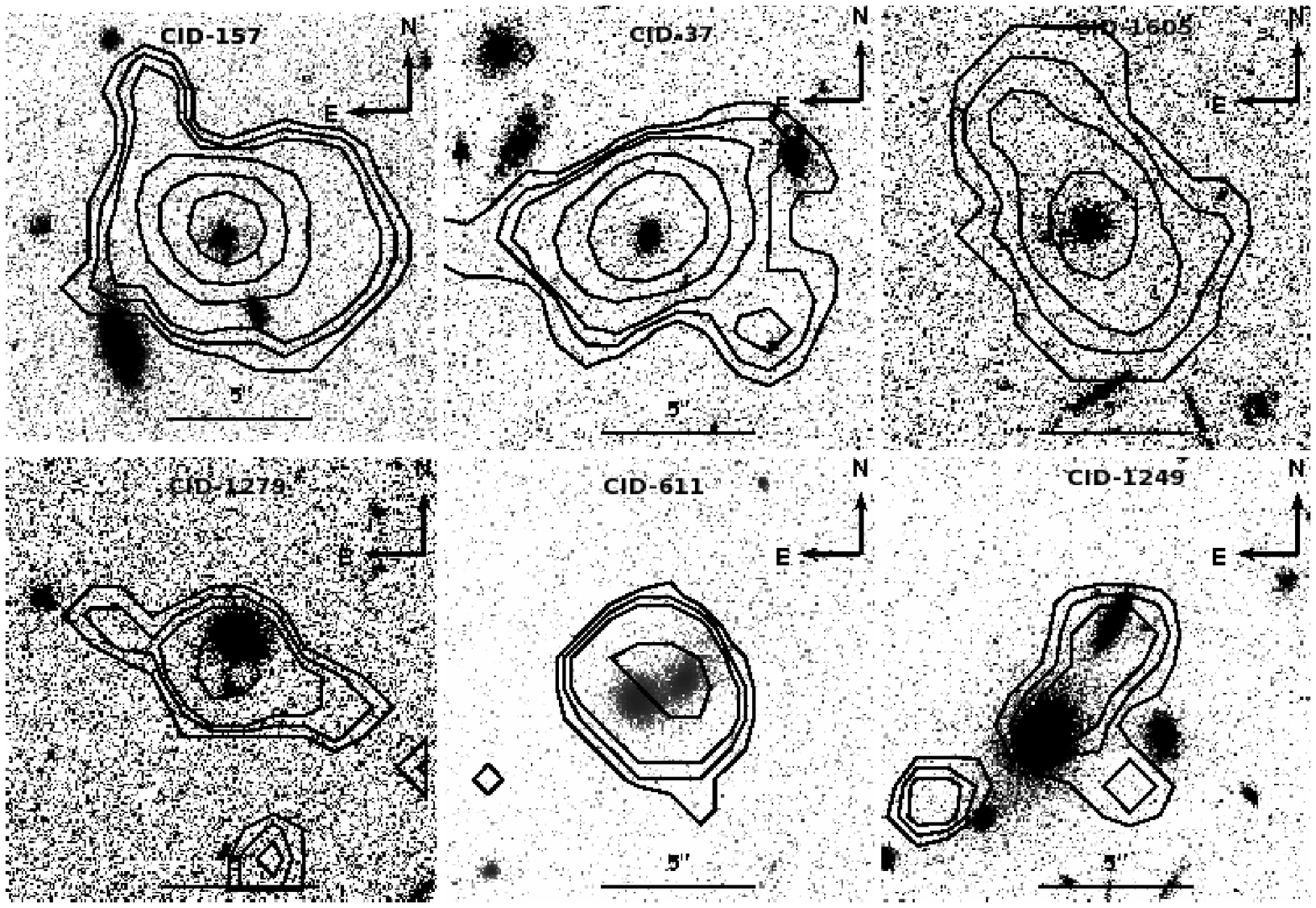}
\includegraphics[width=0.9\textwidth]{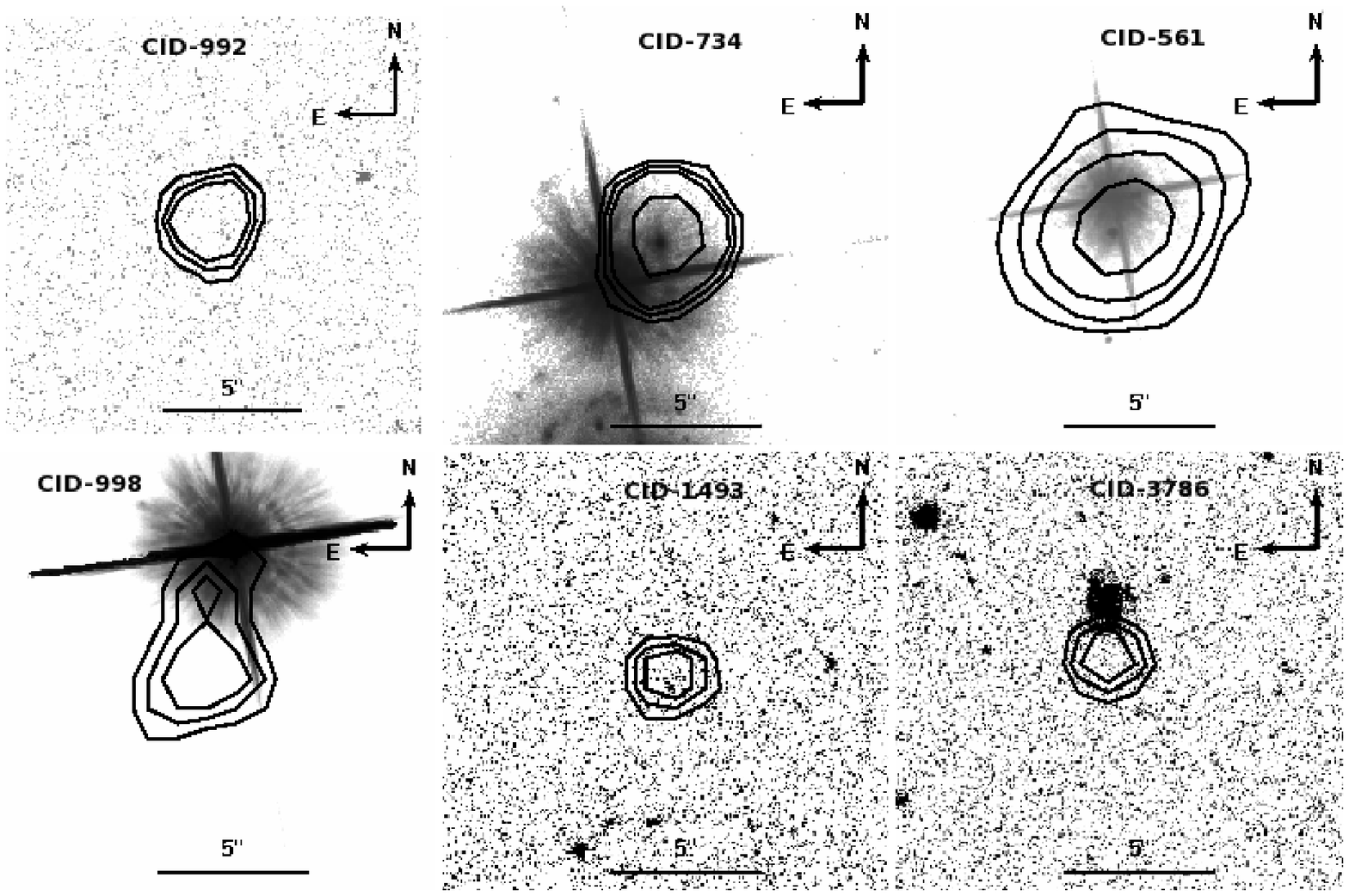}
 \caption{\small Examples of identifications. X-ray contours are overlaid on the ACS images. 
 {\em Secure} identified sources (1 to 3): CID-157, CID-37, CID-1605; {\em Ambiguous} sources (4 to 6): CID-1279, CID-611, CID-1249. 
 The two possible \cps\ are visible in the inner X-ray contour; {\em Unidentified} sources (7-9): CID-992 is a truly empty field at all wavelengths, CID-734 and CID-561 
 are point-like bright sources very close to a bright star so it is not possible to derive their photometry. 
 {\em Sub-threshold} sources (10-12): CID-998 is a bright star at large distance from the X-ray centroid, CID-1493 is a very faint source ({\it i$_{AB}$}=26.43), 
 CID-3786 is a bright optical source but not very close to the X-ray centroid. } 
 \label{ex_fc}
\end{figure*}

Following the same approach used for XMM-COSMOS identifications (Brusa et al. 2007, 2010), the X-ray sources have 
been divided into 4 different classes (see Table \ref{tablike}): 
\begin{enumerate}
\item {\em secure}: sources for which the likelihood procedure is able to find only one \cp\ with $LR >L_{\rm th}$:  
the vast majority of the sources belong to this class; examples are given in panels 1 to 3 in Fig. \ref{ex_fc};
\item {\em ambiguous}: sources with more than one \cp\ above the threshold (panels 4 to 6 in Fig. \ref{ex_fc}); 
\item {\em sub-threshold}: sources with no \cp\ with $LR >L_{\rm th}$ but with possible \cp\ with $LR <L_{\rm th}$ 
within 5''(panels 10, 11, 12 in Fig. \ref{ex_fc}); 
\item {\em unidentified}: sources with no \cp\ in each given catalog, even below the 
threshold, within 5$^{\prime\prime}$ from the X-ray centroid (panels 7, 8 and 9 in Fig. \ref{ex_fc}).
\end{enumerate} 
We anticipate here another, small, class (5) {\em retrieved}, which is not a direct output 
of the likelihood procedure, and which has been introduced after the visual inspection and is discussed later (see  $\S$4.3.1).

\subsection{Ambiguous Source Resolution}

In the class of the {\it ambiguous} sources, there are cases with more than 1 possible \cp\ above the $LR$ threshold. 
For example, in the K band, there are 36 X-ray sources with 2 possible \cps, 4 sources with 3 
possible \cps, none with 4 possible \cps\ and 1 source with 5 possible \cps, all of the \cps\ being above $L_{\rm th}$.
Even though the X-ray sources in Elvis et al. (2009) are all unresolved point sources that show no evidence for extended emission, some of the sources 
with multiple \cps\ could be associated with X-ray groups. However this possibility has not been explored and we assumed that all the X-ray sources 
are associated with a single optical/infrared counterpart. 
We also inspected the location of {\it ambiguous} sources on the field, but their size is consistent with a single point-like source.

\begin{figure}
\centering
\includegraphics[width=0.5\textwidth]{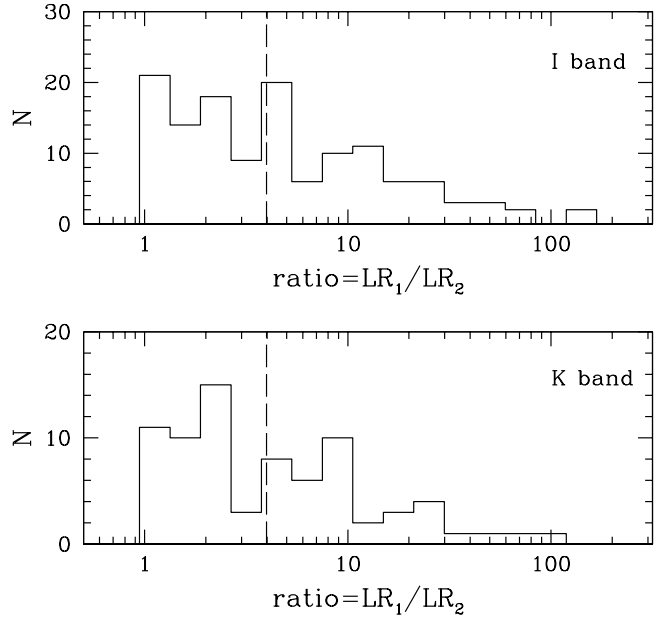}
\caption{\small Histograms of the ratio of $LRs$ of the first and the second possible \cps\ 
for the {\it ambiguous} sources in the {\it i} (top panel) and in the K (bottom panel) band. 
The dashed lines represent the median value of the LR1/LR2 distribution. We adopt this value to solve some of the ambiguities. }

\label{ratio}
\end{figure}

In order to resolve some of the ambiguous cases, we computed the
distributions of the ratios between the $LR$ of the highest $LR$ \cp\ and the other \cps\ for the same X-ray source.
In Figure \ref{ratio} the histogram of the ratio of $LRs$ of the first and the second possible \cps\ is 
reported for the {\it i} and K band {\it ambiguous} sources. A high ratio strongly suggests that the highest likelihood 
counterpart is the correct one. In both {\it i} and K bands,  
if the ratio is above the median value (LR1/LR2=4 in both bands, dashed lines in Figure \ref{ratio}), 
we define the highest $LR$ object as the {\it secure} identification. This method has been applied before in Brusa et al. (2007; private communication). 
Using this criterion, 
by definition, 50\% of the {\it ambiguous} cases in the K and {\it i} band respectively are solved. 
The sources with two possible counterparts are flagged accordingly in the catalog in the identification flag column (see Section \ref{catalog}).
The final numbers of {\it ambiguous} sources in each band after this resolution are reported in Table \ref{tablike}.
\\

\subsection{Identification Rates}
\label{ident_rates}
In Table \ref{tablike} (columns 1, 3 and 5), we report the fraction of sources in each of the classes described
above for each band obtained from the output of the LR analysis.

The K-band catalog identification returns
a higher identification rate of {\it secure} sources than the {\it i}-band identification (90\% vs 85\%), 
fewer sources below $L_{\rm th}$ (4.2\% vs. 9.2\%) and fewer {\it ambiguous} sources (2.7\% vs. 5\%).
The differences are due to the different depths of the {\it i} and K band catalogs. The K-band catalog 
is deep enough to recover most of the \cps, showing the good coupling of X-ray and K 
fluxes, without being too deep to introduce ambiguities and increase the number of spurious associations. 
For comparison, at similar depth both in the optical and in the X-ray, 82\% of sources have been successfully 
identified in the AEGISX survey (Laird et al. 2009), 
while in the CDFS, at fainter fluxes, the fraction of {\it secure} identifications in both the 
optical and K band are slightly lower (74\% and 85\%, Luo et al. 2010).

The 3.6$\mu m$ identification rate is the highest of the three (95\%). 
The depths of the COSMOS 3.6$\mu m$ and K-band catalogs are nearly the same, but 3.6$\mu m$ favors the detection of red objects even more 
than the K band and thus retrieves those undetected in the optical, leading to the increased identification rate. 
Similar 3.6$\mu m$ identification rates (98\% and 94\%) have been found for the ECDFS and the AEGIS-X samples (Cardamone et al. 2008 and Laird et al. 2009) 
at the same X-ray flux limit, but at a magnitude limit of 24.4 AB (5$\sigma$) and 23.8 AB, respectively. 
A somewhat lower identification rate (89\%) has been found in the CDFS (2Ms catalog, Luo et al. 2010) at similar IRAC depth. 

However, the blending and confusion of IRAC sources, due to the larger PSF (1.7'' at 3.6$\mu m$), might contribute to spuriously increasing 
the identification rate. To quantify this effect, we ran the likelihood procedure assuming the same 
background sources distribution used above, the same $L_{\rm th}$ (this is a conservative choice given that a fainter sample implies a lower $L_{\rm th}$ 
to maximize completeness and reliability) and decreasing the flux of the objects with 5'' from the X-ray position by a factor of 2. 
The number of {\it sub-threshold} sources increases to 5.5\%, which is still a low number and comparable 
to the results obtained in the K band. We thus believe that, even if all the 3.6$\mu m$ sources in the sample are contaminated by bright neighbors, and their 
flux is somehow fainter, the secure sources fraction would be still of the order of $\sim$90\%. 
According to this result, the blending problem, which might be still important for SED fitting, 
does not invalidate the results of the likelihood ratio analysis.

 Among the sources with $LR >L_{\rm th}$ ({\it secure} and {\it ambiguous}), we estimated the number of spurious associations in each band. 
This number corresponds to the difference between the number of X-ray sources with at least 1 identification with $LR >L_{\rm th}$ and the expected 
number of real identifications obtained adding the reliability of all the objects with $LR >L_{\rm th}$. 
The obtained percentages are 6\% in the {\it i}-band, 1\% in the K-band and 2\% at 3.6$\mu m$.   

\begin{table*}
\caption{Identification percentages in each band before (columns 1, 3, 5) and after (columns 2, 4, 6) 
the match between the three catalogs, the visual checks and after recovering the faint sources. The final percentages are reported in column 7.}
\begin{center}
\begin{tabular}{clcccccccc}
\hline
\hline
                     && {\it i}  & {\it i}   &  K   &K    & 3.6$\mu m$         & 3.6$\mu m$         &Total \% & Total Number\\
                    && before  & after   & before  & after   &before  & after   & & \\
\hline	
1 & secure id.	      &84.8\% &84.1\% 		     &90.1\% &89.7\% 	  & 95.6\%	  &94.8\% &	  96.3\%  & 1708\\
2 & ambiguous id.     &3.7\%  &3.7\%  		     &2.7\%  &2.6\%  	  & 1.3\%	&1\%	&	2\% & 24 \\
3 & sub-threshold id. &9.2\%  &6.1\% 		     &4.2\%  &2.3\%  	   &1.8\%	&0.8\%  &	1.1\%  &  19\\
4 & unidentified      &1\%    &1.8\%  		     &3.1\%  &3.6\%  	  &1.3\%	&3.4\%  &	0.6\%  & 10\\
5 & retrieved	      &       &3.3\%  		     &       &1.8\%  	  &		&	&		& \\
\hline
\hline
\end{tabular}
\end{center}
\label{tablike}
\end{table*} 

\subsection{Cross-Wavelength Matching of Identifications}

The identifications from the 3 catalogs have been compared in order
to check the consistency among the \cps. 
We performed a positional cross-correlation first between the K and {\it i}-band, and then the K and 3.6$\mu m$ results.  
While checking the results of the cross-correlation, we visually inspected 
the associations by using HST/ACS (filter F814W), K, IRAC (3.6$\mu$m) and \chandra\ 15$^{\prime\prime}\times 15^{\prime\prime}$ 
cutouts around each X-ray source. The ACS cutouts have been used because of the better, subarcsecond, PSF 
with respect to the ground-based {\it i} band images. We also made use of X-ray contour levels, produced with the 
CIAO tool {\sc dmcontour}\footnote{http://cxc.harvard.edu/ciao/ahelp/dmcontour.html} on the \chandra\ mosaic (as shown in Fig. \ref{ex_fc}). 

First, the K and {\it i}-band results have been cross-matched using a radius of 0.7$^{\prime\prime}$, which maximizes the number 
of associations without introducing unreliable matches: 88\% of the X-ray sources 
are associated with the same \cp\ (86.4\% above the threshold in each band), 
while $\sim$7\% of the sources have different associations in the two bands. 

In most of the latter cases, the {\it i} band associations are {\it sub-threshold} while the K band are not, suggesting a 
misidentification in the {\it i} band.  By visually inspecting the cutouts, we verified that in almost all these cases the sources associated with the 
K-band \cps\ are not present in the input optical catalog because of their faintness or low S/N. 
Therefore, the association with the K-band \cps\ has been taken as correct and the initial {\it i} band association discarded. 
We retrieved the {\it i} magnitudes of the K-band \cp\ from the original optical catalog (with no cut in S/N) when available. 

This process leads to a decrease of the number of the {\it sub-threshold} sources in the {\it i} band and the definition of a new class (5)- '{\it retrieved}'. 
When the different associations in the two bands were both secure, we used the 3.6$\mu m$ associations, the cutouts and the X-ray contour levels to help the identification and 
in most of the cases ($>$90\%) the IRAC 3.6$\mu m$ band association agreed with the K-band \cp, strengthening the identification. 

Of the remaining $\sim$5\% of the sources, 2\% are {\it ambiguous} objects in both bands with a separation between the
two possible \cps\ smaller than the \chandra\ PSF at the source position, while 3\% are objects {\it unidentified} in one or both bands. 

Second, the K and 3.6$\mu m$ results have been matched using the same 0.7$^{\prime\prime}$ radius. A good agreement is found: 
96\% of the sources with a \cp\ in both bands are associated with the same source. Among the sources for which we do not find a common \cp\ between the two bands we find 
the same behaviors as above. There are objects classified as {\it sub-threshold} in the K-band and {\it secure} at 3.6$\mu m$ which are replaced 
by faint K-band sources at the 3.6$\mu m$ positions. These sources have been found in the original K-band catalog and included as {\it retrieved} 
in the identification list. The original {\it sub-threshold} K-band \cp\ has been then discarded. 
There are also sources for which a {\it secure} association was found in both bands but these don't coincide. We visually checked 
the cutouts for these sources and we decided which \cp\ to prefer with the help of the {\it i} band associations and the X-ray contours levels.

We then performed a visual inspection of all the sources, according to the different identification classes. At this stage the 3.6$\mu m$ {\it secure} association 
fraction decreases in favor of the {\it unidentified} sources because some of them ($<$1\%), due to the larger PSF, were wrongly 
associated with bright close-by sources (stars or galaxies) and thus are not real 
\cps\ associated with the {\it i} and K-band sources. 
  
At this stage, we also used the revised version of the optical catalog, as explained in Section 2. Among other improvements, this new version of the catalog 
accounts for correctly deblended faint sources close to bright/saturated ones 
(see more details in the changes in the header of the photometric catalog of Ilbert et al. 2009\footnote{Available at http://irsa.ipac.caltech.edu/data/COSMOS/tables/redshift/.}).

\subsection{Summary of Identifications}

The fraction of sources in the three bands for each class, after correlating the $LR$ outputs and the visual checks, 
are reported in Table \ref{tablike} (columns 2, 4, 6). 
We also report the percentage of \cps\ {\em retrieved} in the K-band and {\it i}-band catalogs with a 
magnitude below the formal 5 $\sigma$ limit of the input catalogs.

We assigned a final identification flag that summarizes the results in the 3 bands. The final percentages and total numbers per class are reported in the last two 
columns of Table \ref{tablike}. 
 We define: \begin{enumerate} \item\ {\it secure} identification if the source is a secure identification in at least one band (1708 sources); 
\item {\it ambiguous} identification if the source is an ambiguous identification in at least 2 bands (24 sources); 
\item {\it sub-threshold} identification if the source is in this class in all 3 bands (19 sources); \item {\it unidentified} sources if not identified 
in any of the 3 bands (10 sources). \end{enumerate} 
With the above definitions we have $\sim$97\% {\it secure} identifications, 91\% being 
secure in at least two bands. 

The sources classified as {\it ambiguous} in the final identification are all sources which remain classified as such in both the {\it i} and K bands and 
in all the cases the same source was chosen as the primary solution in both bands.

\begin{figure*}
\centering
\fbox{\includegraphics[width=\textwidth]{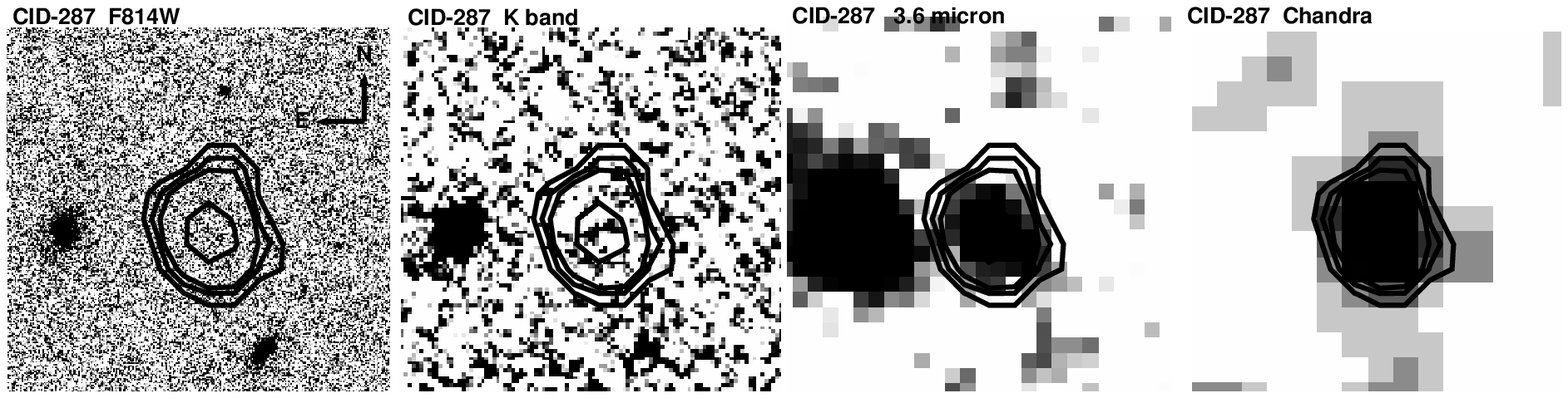}}
\fbox{\includegraphics[width=\textwidth]{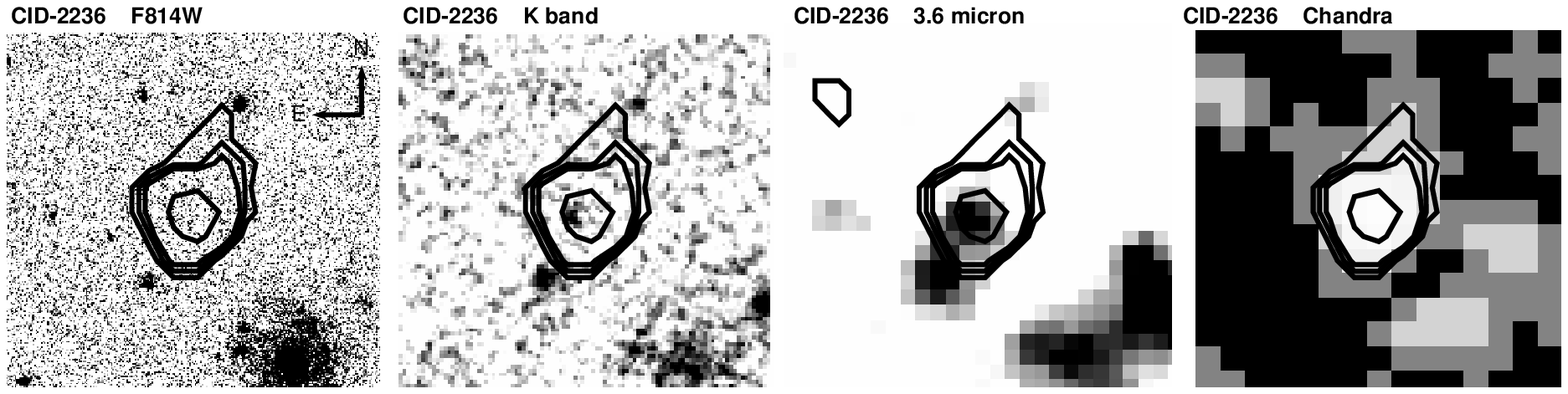}}
 \caption{\small Examples of two sources (CID-287 and CID-2236) identified in the K and 3.6$\mu m$ bands but not in {\it i} band. ACS, K, 3.6$\mu m$, \chandra\ 
 cutouts are shown with overlaid the X-ray contours. Both sources are very faint also in the K band. }
\label{ex_fc_faint}
\end{figure*}

Quite interestingly, there are 19 sources without an optical \cp\ at the X-ray position (see Fig. \ref{ex_fc_faint}). 
Eleven of these have a K and IRAC \cp. Two sources have only a K band \cp, but one of them (CID-425) 
has an optical source visible in the HST image, where however the photometry 
has been totally contaminated by a bright nearby source. Six sources have only an IRAC \cp. 

Ten objects remain {\it unidentified} in all 3 bands. Only two of them are truly empty fields (CID-992, panel 7 in Fig.\ref{ex_fc}, and CID-22962). 
The other eight sources belong to 
the {\it unidentified} class, however, a possible \cp\ exists but is close to bright stars or galaxies, for which reliable photometry is not possible and there is not 
a entry in any COSMOS photometric catalog. Two examples (CID-734 and CID-561) are shown in Panel 8 and 9 of Figure \ref{ex_fc}. These sources have been first associated by the likelihood method 
with the bright stars or galaxies nearby, which have been discarded after visual inspection. The X-ray centroid and contours are clearly pointing to the 
source missing from the catalog. 

In summary, 20 sources do not have an optical \cp\ listed in the catalog and also are truly empty field in the optical. 
In X-ray selected samples, non--detection in the optical band 
has been often assumed to be a proxy for high redshift, or for high obscuration level, or a combination of both, and 
such sources have been dubbed in the past Extreme X-ray Objects (EXOs; e.g. Koekemoer et al. 2004).  
In this respect, only one of the sources with K and IRAC identification has been spectroscopically observed (CID-472) and has been confirmed to be at 
high redshift (z$_{spec}$=3.15). These optically unidentified sources have been included in the high-redshift space 
density computation presented in Civano et al. (2011).

The number of objects detected in all combinations of 3, 2 or 1 bands are reported in Table
\ref{tabcol}. The sources with no detection in the optical should be the reddest objects
($R-K>$5, sources with K and IRAC 3.6$\mu m$, or only IRAC 3.6$\mu m$ as in the examples in Fig. \ref{ex_fc_faint}). 
The objects detected in {\it i} and IRAC bands but not in K are in most cases close to bright sources (K$<$18), so their photometry 
was not extracted in the K-band catalog (McCracken et al. 2010).

To estimate the total number of spurious associations in the field, the number of expected spurious associations among the {\it secure} 
sources (explained in Section 4.2) can be added to the number of sources with $LR <L_{\rm th}$, 
many of which are indeed expected to be either spurious associations or associated to spurious X-ray sources (see Section 5).
However, since we used the combined information from the three bands and performed a visual inspection to isolate possible wrong matches, 
so we expect that the final number of spurious associations will be lower. In conclusion. we can safely assume that 
it is $<$ 6\%, which is in good agreement with the finding in the CDFS of Luo et al. (2010).

Finally, there are also 7 X-ray sources (included in the {\it secure} sources) that are close to bright galaxies, for which the X-ray contours are not centered on the galaxy 
itself but are clearly offset toward the outer region of the galaxy. These `off-nuclear' sources have been 
discussed by Mainieri et al. (2010) and are candidates to be Ultra-Luminous X-ray sources (ULXs).

\begin{table}
\footnotesize
\caption{Number of objects detected in all combinations of 3, 2 or 1 bands and percentage. }
\begin{center}
\begin{tabular}{lcc}
\hline
\hline
&   Number of sources & \% \\
\hline	
{\it i} + K + 3.6$\mu m$	&1652		& 93.9\\
K + 3.6$\mu m$			&	11	 & 0.6\\
{\it i} + K			&	33 	& 1.9\\
{\it i} + 3.6$\mu m$ 		& 32		 &1.8\\
{\it i} 			&	15 & 0.8\\
K				&	2 &0.1\\
3.6$\mu m$			&	6 & 0.3 \\
unidentified		&	10	&0.6\\
\hline
\hline
\end{tabular}
\end{center}
\label{tabcol}
\end{table}

\section{Positional Offsets and Magnitude Distributions}
\begin{figure}[t]
\centering
\includegraphics[width=0.5\textwidth]{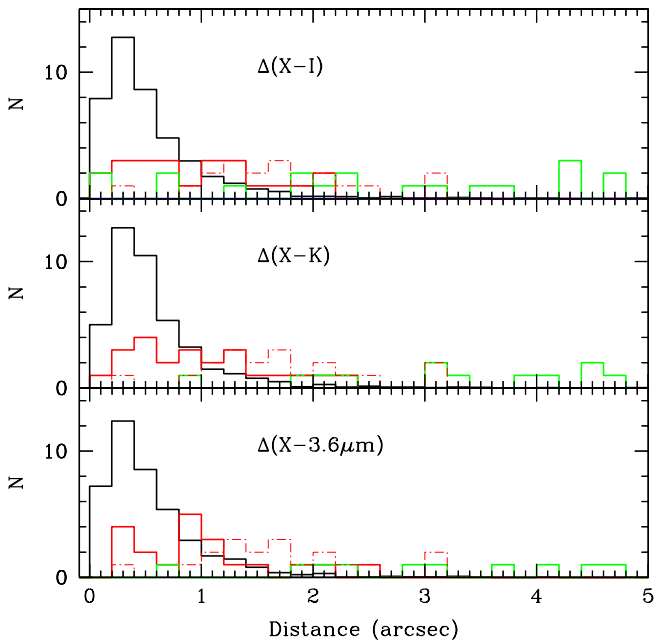}
\caption{\small Histogram of the distances between the X-ray and {\it i} (top), K (middle) and IRAC 3.6$\mu m$
(bottom) \cps\ of the X-ray sources; black: {\it secure} sources (the
histogram is divided by 40 for plotting purposes); red: {\it ambiguous} sources 
(thick: highest $LR$; dashed-thin: second possible); green: {\it sub-threshold} sources; blue: faint {\it retrieved} sources (histogram is divided by 2 for plotting).}
\label{distance}
\end{figure}

The distributions of the X-ray to K, {\it i} and 3.6$\mu m$ positional offsets are shown in Figure \ref{distance}. The
different colors of the histograms refer to the classes of identification ({\it secure}, black; {\it ambiguous}, red; 
{\it sub-threshold}, green) as in the final classification (last column of Table \ref{tablike}). 
Almost 90\% of the securely identified sources in each band lie within 1$^{\prime\prime}$ of the X-ray centroids,
which is consistent with the results obtained by other \chandra\ surveys (e.g., Brand et al. 2006, Laird et al. 2009, 
Green et al. 2009, Luo et al. 2010, Xue et al. 2011, Goulding et al. 2012). This result is 
comparable to the XMM-COSMOS sample distance distribution (90\% within 3$^{\prime\prime}$, Brusa et al. 2007) taking into account 
the larger \xmm\ PSF. 
If the most likely offset between an X-ray source and its counterpart is 
actually at 1$\sigma$ separation, according to the Rayleigh distribution, and given that the positional error ranges 
between 0.2''-1.5'' (from on-axis to off-axis sources; Figure 14 of Paper II) our results are in agreement with 
this expectation.
The objects with offsets larger than 3.5$^{\prime\prime}$ are mostly associated with low signal-to-noise ratio 
X-ray sources (with less than 20 counts in the 0.5-8 keV band) which have larger than average 
uncertainties on their X-ray position ($>$1''). Two of them are associated with off-nuclear X-ray sources (Mainieri et al. 2010).

For {\it ambiguous} sources, the histograms of both the first (red solid line) and the second (red dashed line) possible \cps\  
are plotted. Both \cps\ are close to the X-ray centroid (see Panel 4 to 6 in Fig. \ref{ex_fc}), but 
the first possible \cps\ have a distribution which peaks at slightly lower offsets 
than the second. Only few {\it ambiguous} sources with offsets larger than 2$^{\prime\prime}$ are left after 
cross-correlating the 3 bands. The cross-correlation of the 3 catalogs reduces by half, with respect to 
the numbers in each band, the fraction of {\it ambiguous} sources in the final classification. The final percentage is only 2\%. 

The histogram of offsets for {\it retrieved} \cps\ (class 5, blue histogram) corresponding to columns 2 and 4 of Table \ref{tablike} 
has been reported only for comparison with the {\it secure} identification histograms. The blue histogram follows that of 
the securely identified \cps, suggesting that they are as reliable as the brighter \cps.

\begin{figure}
\centering
\includegraphics[width=0.49\textwidth]{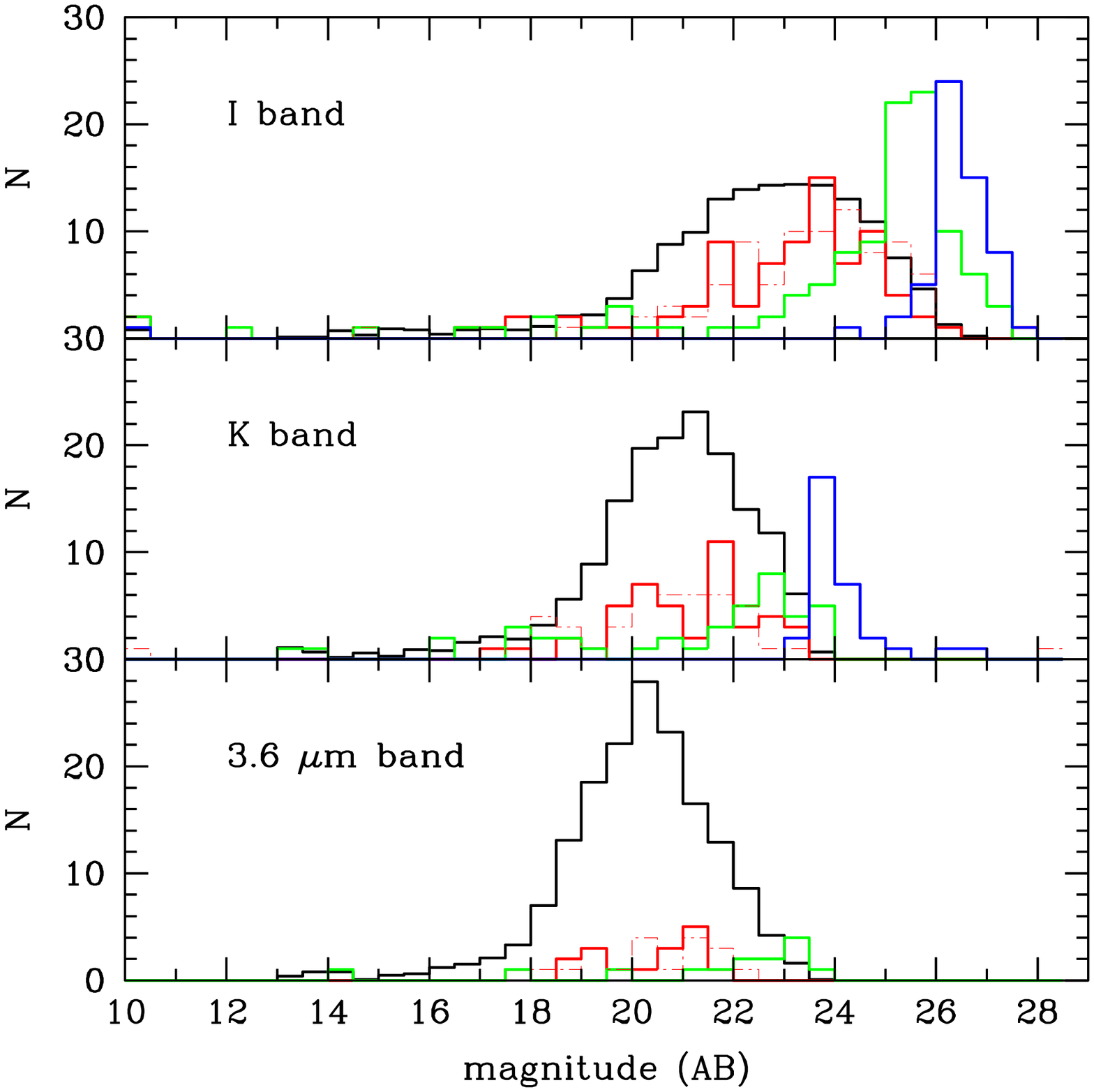}
\includegraphics[width=0.49\textwidth]{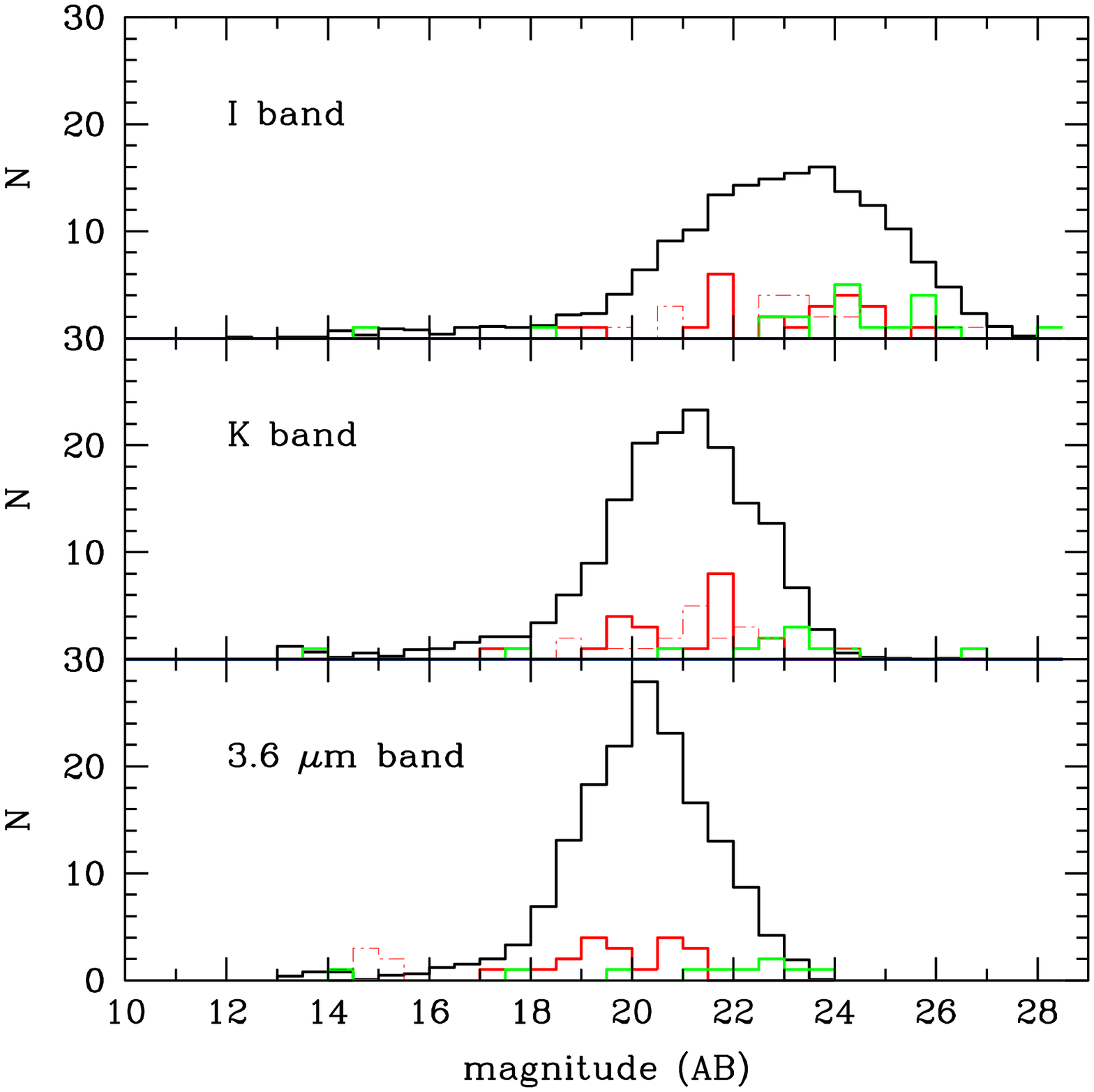}
 \caption{\small Distribution of the {\it i} (top), K (middle), 3.6$\mu$ m (bottom) magnitudes. 
Black: {\it secure} sources (the histogram is divided by 10 for plotting purposes); red: {\it ambiguous} sources 
(thick: highest $LR$; dashed-thin: second possible); green: {\it sub-threshold} sources; blue: faint {\it retrieved} sources. 
 {\it Left:} The magnitude distribution according to the classification before the merging of the 3 bands, corresponding to the 
 columns 2, 4 and 6 in Table \ref{tablike}.
 {\it Right} The magnitude distribution according to the final classification, after merging the results of the 3 bands, as in 
 the last column of Table \ref{tablike}.}
\label{dmag}
\end{figure}

In Figure \ref{dmag}, the magnitude\footnote{The magnitudes plotted are from aperture photometry: 3$^{\prime\prime}$ in the optical, 3$^{\prime\prime}$ in the K band and 
3.8$^{\prime\prime}$ in the IRAC band.} distributions of the X-ray source \cps\ in {\it i} (top), K (middle) and
3.6$\mu m$ (bottom) bands are shown before (left) and after (right) the merging of the classification in each band.  
Thus, in the left panel of Figure \ref{dmag}, we have 5 different classes (corresponding to columns 2, 4 and 6 of Table \ref{tablike}), 
while in the right one only 4 classes (corresponding to column 7 of Table \ref{tablike}). 

The histograms for both \cps\ of the {\it ambiguous} sources (red solid and dot-dashed) follow the shape of the {\it secure} 
source histogram (black) and cover the same magnitude interval; their difference in LR is mainly due to the different 
positional offsets from the X-ray centroid. The primary counterpart (solid histograms in Figure \ref{distance}) is closer to the X-ray position, 
while the secondary is slightly further away from the X-ray centroid.
 
The {\it retrieved} sources (blue histogram in Figure \ref{dmag}, left) lie within one magnitude of the limit of the 
histogram of the {\it secure} sources and, given that they do not show a significantly different distribution of 
X-ray to optical offsets, we consider them as being the fainter tail of the {\it secure} sources. Their contribution is visible 
in the final black histograms of the right panel of Figure \ref{dmag}. 
In contrast, the {\it sub-threshold} objects (green histograms) show significantly larger 
X-ray to optical offsets and, on average, fainter magnitudes than the {\it secure} sources, 
so their low $LR$ values are due to a combination of both factors. 
Six of the 19 {\it sub-threshold} objects are detected in only 1 X-ray band with a low detection probability, 
and given the fact that 5, 4, and 3 spurious detections
with more than 7 counts are expected in the full, soft, and hard bands, respectively, 
they could be associated with spurious X-ray sources.

\section{Spectroscopic and Photometric Redshift Distribution}
\label{Redshiftdistribution}

The coordinates\footnote{We used the optical coordinates for most sources, or
the K-band coordinates if there's no optical \cp, or 3.6$\mu m$ coordinates if there's no K or {\it i} \cp.} of each \cp\ have been 
cross-correlated with the spectroscopic catalogs currently available for the COSMOS field consisting of either data already present in the literature 
(SDSS DR7\footnote{http://www.sdss.org/dr7/start/aboutdr7.html} and also Prescott et al. 2006) or data from the dedicated spectroscopic campaigns: 
 the bright surveys, limited to $i<22.5$, with IMACS at Magellan, Hectospec at MMT (Trump et al. 2007, 2009), 
and VIMOS at VLT (zCOSMOS bright; Lilly et al. 2007, 2009); the DEIMOS survey at Keck, limited to $i<$25, as the result of a multi-year observing campaign (PIs: 
Capak, Kartaltepe, Salvato, Sanders, Scoville; see Kartaltepe et al. 2010); the zCOSMOS deep survey to B$<$25 (Lilly et al. 2009); 
the star survey with the FAST spectrograph on the
FLWO 1.5m telescope (Wright, Drake and Civano 2010).

We also retrieved the WFC3 imaging and grism data of the COSMOS field (van Dokkum et al. 2011) from the archive 
and processed them with standard software. The detection F140W image was reduced using the STSDAS \texttt{pyraf} task 
\texttt{multidrizle} (Koekemoer et al. 2002). The dispersed data (G141 grism) were reduced using the 
\texttt{aXe} slitless spectroscopy package Kummel et al. (2009). The extracted 2D and 1D spectra were visually 
inspected and contaminated sources were rejected. From the set of 19 acceptable spectra, we determined new redshifts 
for 4 sources as well as confirmed 15 redshifts based on ground-based spectroscopy.

At present, 1069 objects (60\%) have spectroscopic redshifts (with 1 or more emission or absorption lines); 906 (51\% of the total) of these have a quality 
flag (3 or 4, see Section 7) corresponding respectively to a secure redshift with 2 or more emission or absorption lines and a secure redshift with 
2 or more emission or absorption lines with a good-quality, high signal-to-noise ratio spectrum 
(see Lilly et al. 2007, 2009 for thorough explanation of quality flags). 

The available spectra are primarily the results of spectroscopic campaigns focused on XMM-COSMOS sources (Brusa et al. 2010).
The spectroscopic campaigns targeting \chandra-only sources started recently, 
when the C-COSMOS \cps\ have been targeted by the deep zCOSMOS survey (Lilly et al. 2009) and by 
the DEIMOS survey. 
For these reasons, the fraction of sources that are not XMM detected, yet have a spectroscopic redshift, is only 35\%. 
In the interim, photometric redshifts provide good estimates on the redshift range 
and classifications of the \chandra\ sources without a spectroscopic identification.

At a magnitude limit of {\it i}=22.5 (AB mag), C-COSMOS is 83\% spectroscopically complete. If, at the same magnitude limit, we consider only bright 
X-ray fluxes similar to XMM-COSMOS ($>$10$^{-15}$\cgs\ in the soft band) the survey is 89\% spectroscopically complete, 
as most of the sources are in common with XMM, 
while, at lower X-ray fluxes ($<$10$^{-15}$\cgs), it is 76\% complete. At fainter 
optical magnitudes ($>$22.5), only 31\% of sources have spectroscopic redshifts.

Tuned photometric redshifts for the C-COSMOS sources have been computed and presented in Salvato et al. (2011).  
In that paper, we used the publicly available code LePhare\footnote{www.oamp.fr/people/arnouts/LE\_PHARE.html} (Arnouts et al. 1999, Ilbert et al. 2006), 
which is based on a  $\chi^2$ template-fitting procedure.
Two libraries of templates were used, depending on a prior on morphology, optical variability and X-ray flux of the source. 
The first library (defined in Salvato et al. 2009, Table 2) consists of AGN templates, hybrids (host +AGN) templates, and few normal galaxies. 
It was used for all the point-like sources in the HST images (as defined by Leauthaud et al. 2007) that presented optical variability (see Salvato et al. 2011 for details), 
imposing a luminosity prior on the absolute {\it B} magnitude typical for bright AGN and QSO (-20$<M_B<$-30). The same library was used also for extended sources 
with an X-ray flux brighter than 8$\times$10$^{-15}$ \cgs. In this case, however, no luminosity prior was adopted. 
The second library (as defined in Ilbert et al. 2009) includes only normal galaxies templates and 
it was used for the remaining sources (i.e. extended, non variable and with X-ray flux  $<$8$\times$10$^{-15}$ \cgs), without any luminosity prior. 
The flow-chart in Figure 6 of Salvato et al. (2011) summarizes the procedure. 

In addition to the two above libraries, all the sources have been fit also by various stellar templates (low-mass stars, subdwarf O and B stars, white dwarfs). 
Whenever (2$\times$ $\chi^2_{star}$) $<$ $\chi^2_{gal}$, the source is assumed to be a star (see discussion in Salvato et al. 2011).
For 10 sources, not observed spectroscopically, the photometric fitting gives 
a better $\chi ^2$ with a stellar template, and the visual inspection of the ACS images confirms the stellar nature of these sources.

The total number of sources with a photometric redshift is 1693. 
The accuracy achieved is $\sigma_{\Delta z}/ (1+z_{spec}) \sim$1.5\% (with 5.8\% outliers) on the whole sample with no cut in magnitude or redshift applied.
These results are comparable in precision to those achieved typically for non-active galaxies (Cardamone et al.  2008, Wuyts et al. 
2008, Ilbert et al. 2009, Barro et al. 2011) and to those reached only recently for photometric redshift of AGN (Salvato et al. 2009, Cardamone et al. 2008, Luo et al. 2010) 
with a similar number of photometric bands.

\begin{figure}
\centering
\includegraphics[width=0.5\textwidth]{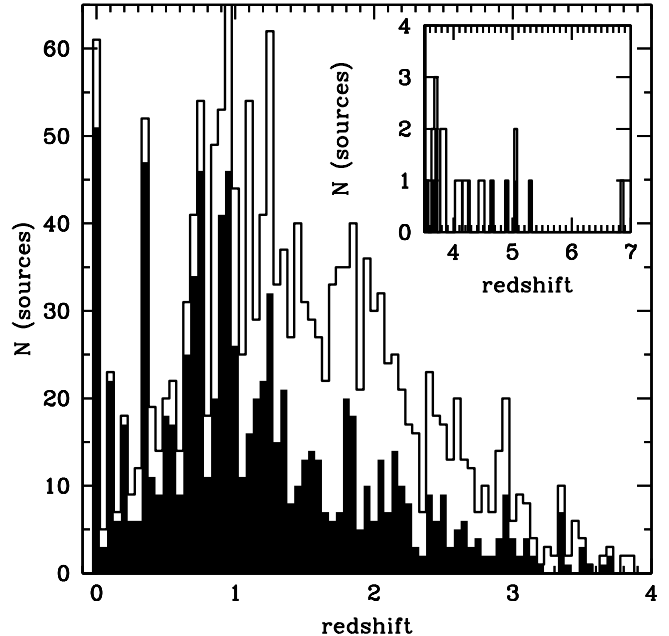}
\caption{\small Redshift distribution of the optical \cps\ ($\Delta z$=0.05): open histogram = all sources 
for which either a spectroscopic or photometric redshift is available;  filled histogram= all sources 
for which a spectroscopic redshifts is measured (906). 
In the small insert, a zoom in the high-z region (z$>$3.5) has been plotted. }
\label{zdist}
\end{figure}

The distribution of all the sources with a spectroscopic (906, solid histogram) or a photometric redshift (1693) is reported in Fig. \ref{zdist}. 
The distribution peaks between z=1--2 and the spikes correspond to well known large scale structures in the COSMOS field (Gilli et al. 2009).
There are 75 sources with redshift $>$ 3. These have been presented in Civano et al. (2011), where the number counts and the space density 
of the high-redshift sample have been computed. 
While a sizable sample of high redshift $z>$5 quasars has been 
collected in optical surveys (Fan et al 2006, Willot et al. 2007), only two $z>$5 X-ray selected AGN have previously 
been reported in the literature: one in the CFDN (Steffen et al. 2004, z=5.4) and one in the CLASXS survey (Barger et al. 2005, z=5.19), both of them 
being unobscured type 1 sources. Given the faint optical magnitude, no sources with spectroscopic redshift at $z>$5 have been found in the CDFS, 
while there are 5 photometric candidates (Luo et al. 2010, Xue et al. 2011). In the C-COSMOS sample there are 4 sources at z$>5$, two of them with spectroscopic 
redshift (one with broad lines, Capak et al. 2011; one with narrow lines, Ikeda et al. 2011) and two with photometric redshift, one of them being possibly 
the highest redshift X-ray selected source at z=6.84 (Salvato et al. 2011).


\begin{table}
\footnotesize
\caption{\small Number of X-ray sources by Spectral or Photometric type.}
\begin{center}
\begin{tabular}{lcc}
\hline
\hline
&   Number of sources  & \% \\
\hline
Spectra &&\\	
\hline
Broad Line	& 316	&	35 \\	  
Not Broad Line	& 534	 & 59	\\  
Star	&51	&5.6 \\
Off nuclear (host galaxy)	&5	& 0.4	\\

\hline
Photo-z &&\\	
\hline
Unobscured AGN template& 450	 &27\\
Obscured AGN template	&104	 &6 \\
Galaxy template	&1101	 &65\\
Star template	& 30	&2\\
\hline
Photometric star & 11  & \\
Off nuclear &	7	& \\
\hline
\hline
\end{tabular}
\end{center}
\label{tabtype}
\end{table} 

\subsection{Spectroscopic and Photometric Classification}

Table \ref{tabtype} breaks down the sources on the basis of their spectral type if spectra are available, or by the type of template which best fit the 
photometry of the sources. 

Of the sources with good quality spectroscopic redshifts,   
35\% show at least one broad (FWHM$>$2000 km/s) emission line in their spectra (BLAGN), while 59\% do not present broad lines (i.e., they show 
narrow emission lines or absorption lines only), so we will refer to them as ``non broad line AGN'' (non-BLAGN). The latter sources have not been classified into 
star-forming galaxies or Type 2 AGN because for most of them either the observed wavelength range or the low S/N do not allow the 
use of standard optical diagnostic diagrams (e.g., Kewley et al. 2001).
The remaining spectroscopic sources are spectroscopically identified stars (51 objects, 6\%).
The stars are mostly late-type K and M stars (35), with a number of solar-type F and G stars (16), that mostly lie at distances between 100 and 1000 pc, suggesting that they are primarily Galactic disk sources. 
A small fraction of these sources with high X-ray luminosities lie at greater distances ($>$ 1 kpc) and 
therefore are probably Galactic halo stars (Wright, Drake and Civano 2010).
Given the un-coherent spectroscopic follow-up, the percentages of the different spectral types are not necessarily representative 
of the whole C-COSMOS sample. 

The classification obtained from the photometric fitting has been compared against the spectral classification for 
the sources with spectroscopic redshift. There is a good overall agreement between the two classifications: 
81\% of BLAGN are fitted with the template of an unobscured AGN, while 92\% of the sources classified as non-BLAGN are 
well fitted either by a galaxy template (78\%) or by an obscured AGN template  (22\%). It is known that BLAGN SEDs, especially the low luminosity ones, 
can be contaminated by stellar light (Luo et al. 2010, Lusso et al. 2010, Elvis et al. 2012 sub.); 
thus it is not surprising that 19\% of BLAGN have a different classification 
than an unobscured source. Given this agreement, we can use either the spectroscopic or the photometric classification.
 
\begin{figure}
\centering
\includegraphics[width=0.5\textwidth]{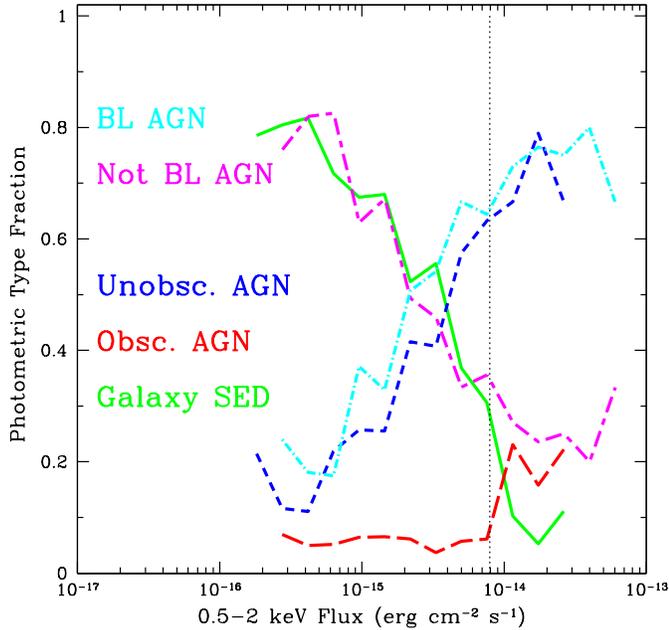}
\caption{\small Fraction of the spectroscopic and photometric classification types of C-COSMOS opt/IR counterparts as function of the soft band X-ray flux. 
Cyan (short - long dashes): BL AGN; Magenta (dot-dashed line): not BL AGN; Blue (short dashes): unobscured AGN; Red (long dashes): obscured AGN; 
Green (solid): galaxies. The vertical dot-dashed line indicates the X-ray flux threshold below which, for extended and optically non varying sources, 
templates of normal galaxies have been used when computing photometric redshifts. }
\label{frac}
\end{figure}

65\% of the C-COSMOS sources are best fit by a non-active galaxy, 
6\% by a template which represents an obscured AGN, and 27\% by a template which includes 
an unobscured AGN component. The fraction of sources by type (both spectroscopic and photometric) as function of the 
X-ray flux are plotted in Fig.~\ref{frac}. 
The number of unobscured or BLAGN decreases with the X-ray flux, while the fraction of galaxy SED sources or non-BLAGN increases 
towards faint X-ray fluxes becoming the dominant population at fluxes fainter than 3$\times$10$^{-15}$ \cgs\ (0.5-2 keV). 
The galaxy SED source and obscured AGN fractions are complementary, because, as stated above, the optically-extended C-COSMOS sources have been 
fitted with a galaxy template library if their X-ray flux is fainter than 8$\times$10$^{-15}$ \cgs\ (0.5-2 keV; vertical line in Fig.~\ref{frac}) and with 
hybrid templates (obscured AGN) if brighter. The small fraction of obscured AGN at fluxes fainter than 8$\times$10$^{-15}$ \cgs\ (where all the extended sources have 
been fit only with galaxies templates) is due to the presence of sources with a point-like morphology, whose best fit is an obscured AGN.

A similar fraction for the galaxy SED objects (48\%) has been found in the photometric classification of CDFS sources at a flux limit of 
1.9$\times$10$^{-17}$ \cgs\ (0.5-2 keV; Luo et al. 2010), while 
a lower fraction of unobscured AGN is found (20\%) and a higher one of obscured AGN (32\%). However, at the fluxes of the CDFS, 
$\sim$35\% of the galaxy SED sources are likely to 
be low luminosity normal and starburst galaxies (Xue et al. 2011, Lehmer et al. 2012).

\begin{figure}[t]
\centering
\includegraphics[width=0.49\textwidth]{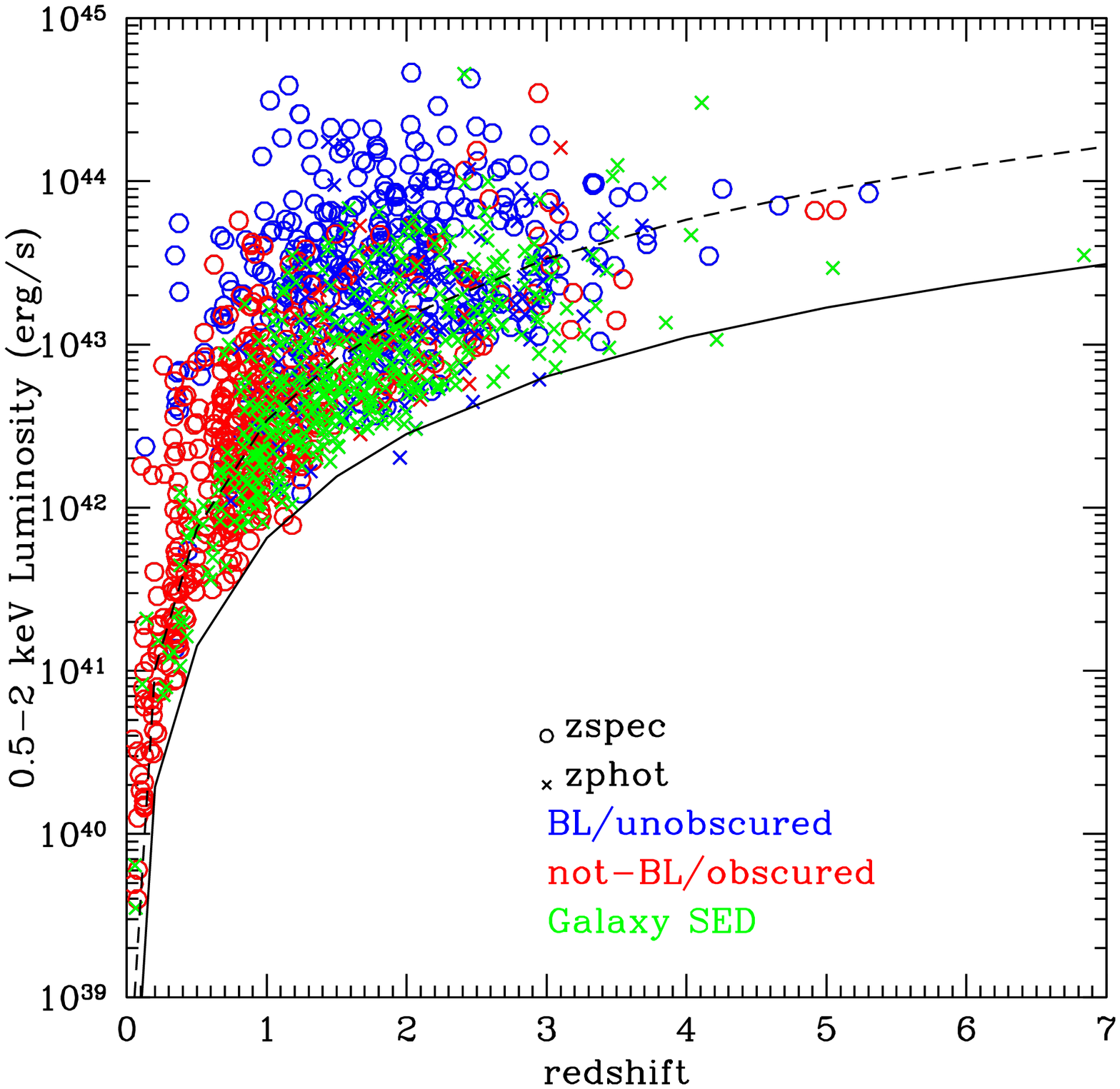}
\includegraphics[width=0.49\textwidth]{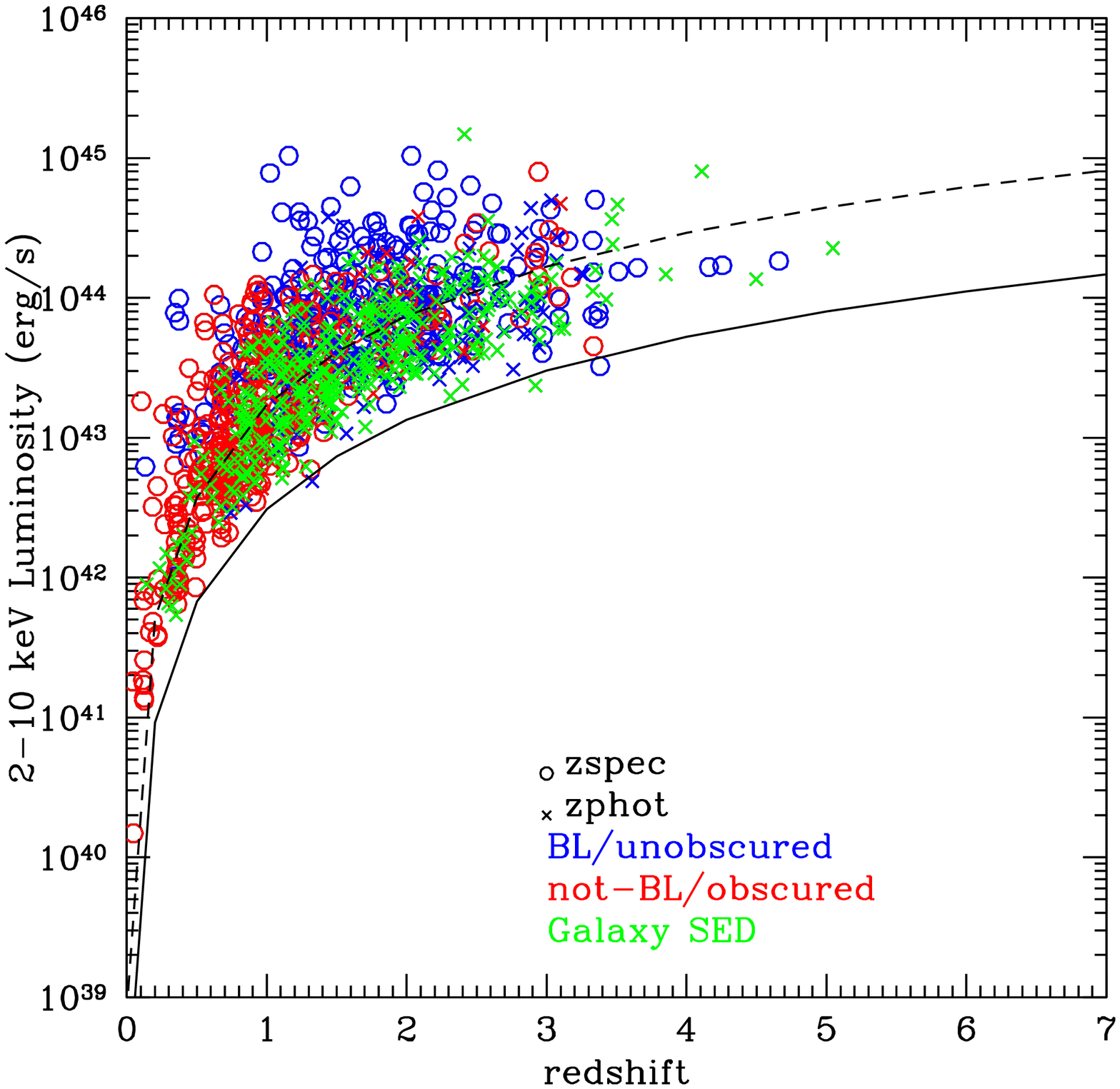}
\caption{\small The luminosity-redshift plane (soft-left, hard-right) for all the sources with spectroscopic or 
photometric redshift, along with their classification. Sources with a spectroscopic identification are plotted as 
open circles, sources with only photometric identification as crosses; the symbols are color coded as follows: blue=BLAGN or unobscured AGN from 
the SED, red=non-BLAGN or obscured AGN from the SED fitting, green= galaxy dominated sources only from the SED. The solid lines represent the C-COSMOS 
flux limit and the dashed lines represent the XMM-COSMOS limit. } 
\label{lz}
\end{figure}

In Figure \ref{lz}, the X-ray luminosity versus redshift plane is shown for the soft (left) and  hard (right) bands.
To convert fluxes into luminosities, both spectroscopic and photometric redshifts were used (giving preference to the 
spectroscopic redshift, when available) and an X-ray spectral index $\Gamma$=1.4 (the same used to compute the X-ray fluxes in Paper I) 
was assumed to take into account the $k$-correction. 
In the Figure we also report the faintest flux limits (dashed line) for XMM-COSMOS from Cappelluti et al. (2009, Table 2).
There are 1323 and 1103 sources in the soft and hard bands, respectively, of which only 12\% and 4\% 
have luminosities possibly not due to nuclear emission or consistent with low luminosity AGN ($< 10^{42}$\lum). The fact that only a small percentage 
of the sources has a luminosity consistent with that of a normal or starburst galaxy, is in agreement with the fact that most of 
the galaxy SED sources (65\% of the total) are candidates to be obscured AGN (as found in previous work, see Mainieri et al. 2005) and not normal galaxies. 

 In the rest of the paper we will use the 0.5-10 keV luminosity to exclude normal and star-forming galaxies (L$_X < 10^{42}$\lum) from the sample.
For 92 sources out of 906 with a spectroscopic redshift, a classification based on standard 
emission line diagnostic diagrams is available from Bongiorno et al. (2010): 66\% of them (61 of 92) have emission line ratios typical of star-forming galaxies, while 
the remaining have the ratios of an obscured AGN. However, only 23\% of those classified as star-forming galaxies (14 out of 61) have a luminosity $< 10^{42}$\lum\ (0.5-10 keV), showing that optical  
 diagnostic diagrams can be insensitive to hybrid objects (obscured AGN with enhanced starformation). A luminosity cut at $10^{42}$\lum\ (0.5-10 keV) may be a more reliable diagnostic to separate 
 star-forming galaxies from AGN, but at the risk of excluding bona fide, low-luminosity AGN.
The coupling of diagnostic diagrams with X-ray luminosities could then be a more effective method to separate obscured AGN from truly 
non-active galaxies (see also Trouille et al. 2011). 

In the 1$<z<$2 redshift bin, C-COSMOS luminosities span about two orders of magnitude (10$^{42.5}$-10$^{44.5}$ \lum). 
C-COSMOS covers the whole redshift range 0 to 3 at X-ray luminosities greater than 10$^{43}$ \lum\ in the soft band and 
10$^{43.5}$ \lum\ in the hard band (solid lines in Fig. \ref{lz}). With this sample, it will possible to derive accurate X-ray luminosity
functions in both the hard and soft bands, in small redshift bins (Miyaji et al. 2012, in prep.).

\section{The Multiwavelength Catalog of C-COSMOS sources}
\label{catalog}
The full multiwavelength catalog of the \chandra\ COSMOS source identifications, together with  multiwavelength cutouts, 
X-ray and optical spectra, is available on the web page http://chandracosmos.cfa.harvard.edu. 
The electronic version of the catalog will contain the multiwavelength properties as listed below. 
Right Ascension and Declination are in the J2000 coordinate system. The catalog will be available with the paper 
in the published version, in the COSMOS repository\footnote{http://irsa.ipac.caltech.edu/data/COSMOS/tables/} 
and also at this link\footnote{https://hea-www.cfa.harvard.edu/$\sim$fcivano/C\_COSMOS\_identification\_catalog.fits}.


\begin{itemize}
\item[-] {\em Column 1:} \chandra\ source name, following the standard IAU
  convention with the prefix ``CXOC'' for '\chandra\ X-ray Observatory COSMOS'
  survey, as in Paper I.
\item[-] {\em Column 2:} Source number. Sources are listed in order of
  detection as in Paper I: first those detected in the full band with $detml \geq$10.8,
  followed by those detected in the soft band only and by those detected in the hard band only.

\item[-] {\em Column 3-4:} The X-ray coordinates of the source from the Paper I catalog.
\item[-] {\em Column 5-6-7:} The X-ray flux in the full, soft and hard bands. Negative fluxes represent upper limits 
as computed following the prescriptions of Paper II.

\item[-] {\em Column 8-9:} The coordinates of the optical/IR \cp, as determined in this paper.
\item[-] {\em Column 10:} The identifier number from the optical catalog of Capak et al. (2007). 
\item[-] {\em Column 11:} The identifier number from the revised version of the optical catalog (Ilbert et al. 2009). 

\item[-] {\em Column 12-13:} The coordinates of the optical \cp\ (Capak et al. 2007, Ilbert et al. 2009). 
\item[-] {\em Column 14-15:} {\it i} band magnitude and error in 3$\prime\prime$ aperture from the optical catalog.

\item[-] {\em Column 16-17:} K-band \cp\ coordinates from the catalog of McCracken et al. (2010).
\item[-] {\em Column 18-19:} K-band aperture magnitude (at 3$\prime\prime$) and error from the catalog of McCracken et al. (2010).

\item[-] {\em Column 20-21} The coordinates of the 3.6$\mu$m \cp\ from the IRAC catalog (Sanders et al. 2007). 
 \item[-] {\em Column 22-23:} 3.6$\mu$m flux ($\mu$Jy) and error in 1.9$\prime\prime$ aperture from the IRAC catalog (Sanders et al. 2007). 
 To convert to total flux, the standard factor suggested in the IRAC user guide has to be applied (division by 0.765).

\item[-] {\em Column 24:} Final identification flag: 1 = {\it secure}, 10 = {\it ambiguous}, -99= {\it unidentified}, 100={\it sub-threshold}.
\item[-] {\em Column 25:} Star Flag to isolate stars: 1 = spectroscopic confirmed star, 
10 = photometric star, 100 = visually identified star. 
\item[-] {\em Column 26:} Off-nuclear flag to isolate the 7 off-nuclear sources from Mainieri et al. (2010).

\item[-] {\em Column 27:} Spectroscopic redshift from the catalogs listed above (Section \ref{Redshiftdistribution}).

\item[-] {\em Column 28:} Spectroscopic identification. The identification flag 
can be read as follows: 1 = BLAGN; 2 = non-BLAGN; 0 = star.       
\item[-] {\em Column 29:} Spectroscopic redshift quality: 3 = 2 or more emission and/or absorption lines; 4 = good signal-to-noise ration and 
2 or more emission and/or absorption lines.
\item[-] {\em Column 30:} Origin of the spectroscopic redshift: 1 = SDSS, 2 = MMT, 3 or 4 = IMACS, 5 = zCOSMOS bright, 6 = zCOSMOS deep, 7 = Keck 
8= other (FAST telescope, FMOS, WFC3).

\item[-] {\em Column 31:} Photometric redshift from Salvato et al. (2011).
\item[-] {\em Column 32:} Photometric identification from the SED fitting (1 = unobscured, 2 = obscured, 3 = galaxy). 
\item[-] {\em Column 33:} The identifier number in the XMM-COSMOS catalog (Cappelluti et al. 2009, Brusa et al. 2010).


\end{itemize}

\section{Optical and Infrared properties}
In this section the basic X-ray to optical properties of the sample are reported. Hereinafter, we will consider only
the 1708 sources with {\it secure} identifications and the \cp\ with the highest $LR$ for the 24 sources with {\it ambiguous} associations. 
We will not consider the 7 off-nuclear sources (Mainieri et al. 2010), the 62 stars  (Wright, Drake and Civano 2010), 
the 10 {\it unidentified} sources and the 19 {\it sub-threshold} sources.

\subsection{X-ray to optical flux ratio}
The X/O ratio (Maccacaro et al. 1988) is defined as 
\begin{equation}
X/O = log(f_X/f_{opt})= log(f_X) +C+m_{opt}/2.5 
\end{equation}
where $f_X$ is the X-ray flux in a given energy range, $m_{opt}$ is the magnitude at the chosen optical wavelength and 
C is a constant which depends on the specific filter used in the optical observations. 
Usually, the {\it r} or {\it i} band flux is used (e.g., Brandt \& Hasinger 2005). 
Originally, a soft X-ray flux was used for this relation (see Introduction), and the majority of luminous spectroscopically identified AGN (both BLAGN and NLAGN) in the Einstein and ASCA surveys 
were characterized by X/O=0$\pm$1 (Schmidt et al. 1998, Stocke et al. 1991, Akiyama et al. 2000, Lehmann et al. 2001). 
With the advent of harder surveys, the same relation has been used in the hard band, without really accounting for the X-ray band 
used or the change in spectral slope (Alexander et al. 2001, Hornschemeir et al. 2001, Fiore et al. 2001, Brusa et al. 2003, Perola et al. 2004, 
Civano et al. 2005, Cocchia et al. 2007, Brusa et al. 2007, Laird et al. 2008, Xue et al. 2011). 

\begin{figure*}
\centering
\includegraphics[width=0.49\textwidth]{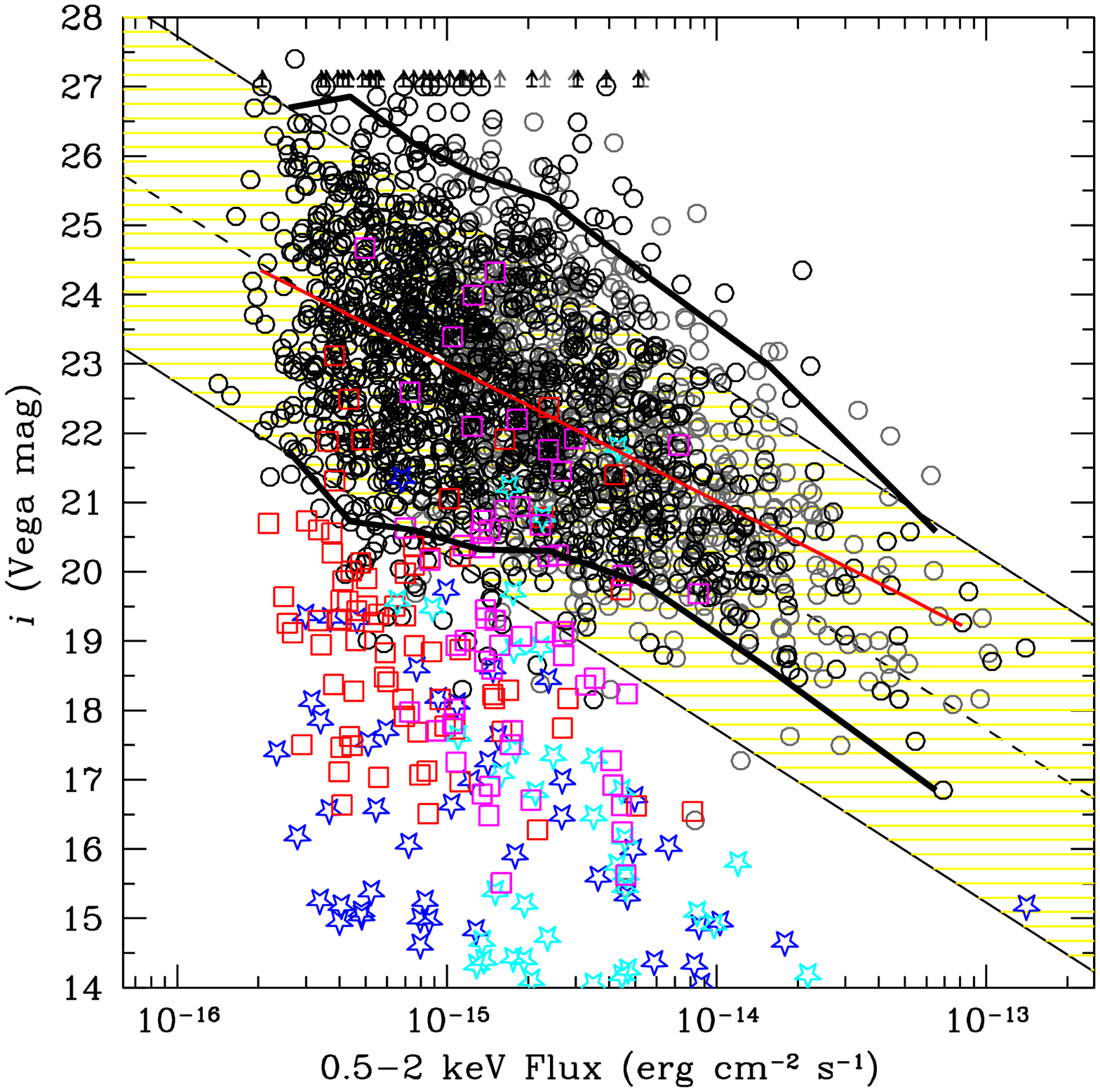}
\includegraphics[width=0.49\textwidth]{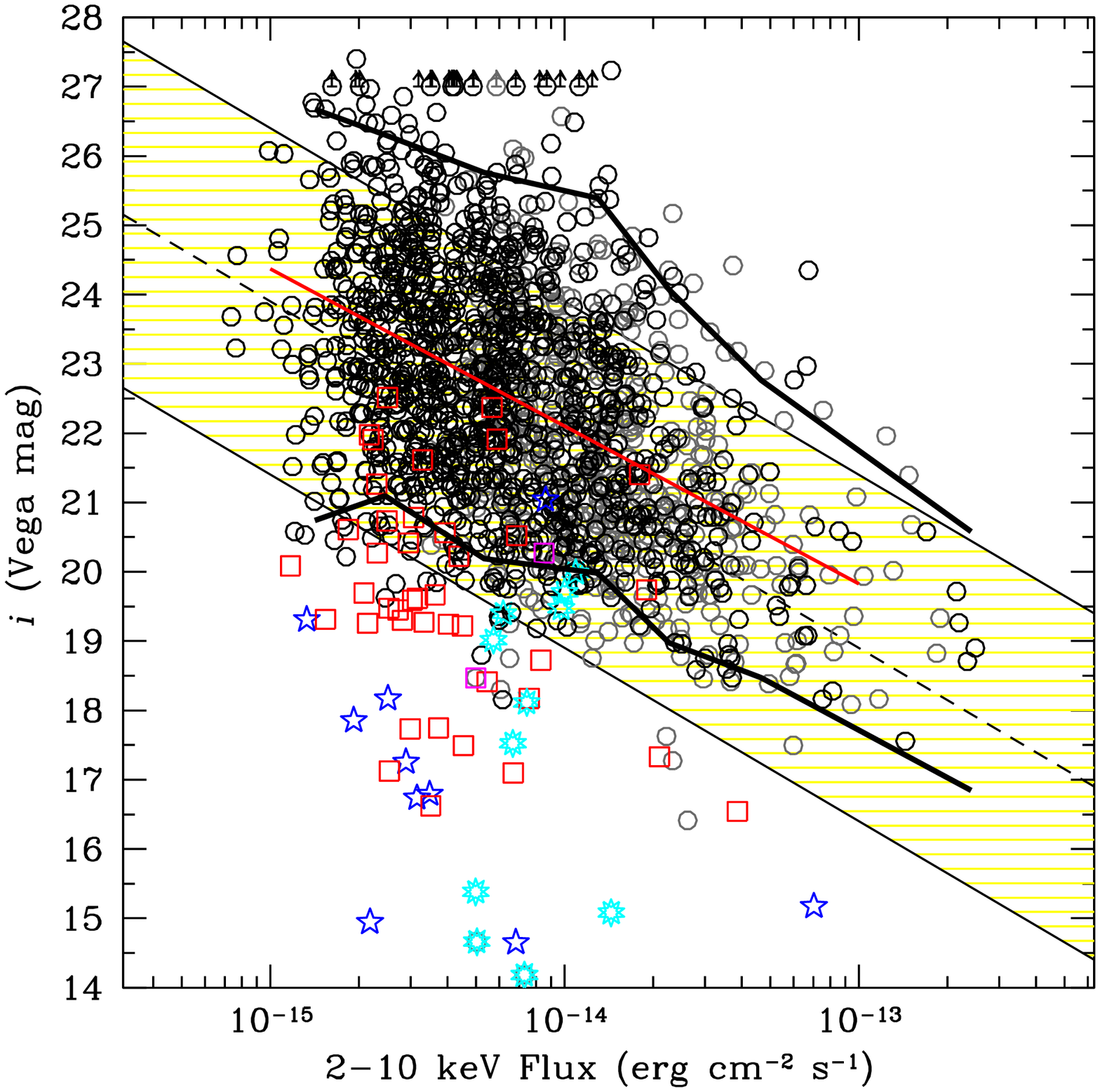}
\caption{\small The X-ray flux (soft-left, hard-right) versus the {\it i}-band total (i.e. aperture corrected) magnitude for all the X-ray sources with a 
{\it i}-band \cp. The yellow shaded region represents the ``classic locus'' and the ``hard band locus'' of AGN along the 
correlation X/O=0$\pm$1. Black symbols: C-COSMOS sources. Gray symbols: XMM-COSMOS sources. Star symbols (blue = \chandra\ and cyan=XMM) represent spectroscopic or 
photometric identified stars. Red and magenta squares represent sources with L$_F<$10$^{42}$ \lum\ in 
C-COSMOS and XMM-COSMOS, respectively. 
Sources not identified in the optical but with a counterpart in the 
K or IRAC band are represented with a upwards arrows. The thick solid curves represent the newly drawn locus. 
The red solid line is the fit to the combined (\chandra\ and XMM) sample, respectively.}
\label{xo}
\end{figure*}

In Figure \ref{xo}, the {\it i} band total magnitude is plotted versus the X-ray soft (left) and hard (right) fluxes for all the sources with 
{\it secure} identification. For both bands, the $X/O=\pm 1$ locus (yellow area) has been defined using as constant $C(i)=$5.91, 
computed taking into account the width of the {\it i}-band filters in COSMOS (Subaru, CFHT or for bright sources SDSS). 
 In the hard band, the locus (hereafter ``hard band locus'') is plotted taking into account the band width and the spectral slope 
used to compute the C-COSMOS fluxes ($\Gamma$=1.4). The difference between the ``classic locus'' and the ``hard band locus'' is X/O=+0.47.    
Sources not identified in the optical, but with a counterpart in the K or IRAC 3.6~$\mu m$ 
band, are represented with upwards arrows at {\it i}=27. The stars in the sample are represented with blue star 
symbols and sources with a full band luminosity L$_{F}<$10$^{42}$\lum\ are represented as red squares. 

The X-ray-vs-optical flux plot is quite homogeneously covered by the C-COSMOS sources with a large 
number of sources outside the X/O=0$\pm$1 locus, mostly in the hard band (Fig. \ref{xo}, right). 
In order to quantify the width of the distribution (i.e. the X/O distributions), we computed the region which includes 90\% of the AGN population (i.e., 
excluding sources identified as stars and sources with L$_{F}<$10$^{42}$\lum) by tracing the 5\% lower percentile 
and 95\% upper percentile of the distributions of i-band magnitudes for sources in X-ray flux bins of width 0.25 dex (black thick solid curves in Figure \ref{xo}). 
In deriving the 95\% upper percentile, we included the non-detections in the optical band. A total of 1274 and 1115 sources were used in the soft and hard band, respectively. 
We refer in the following to this region as newly drawn locus. 

In the soft band, the newly drawn locus fits very well the ``classic locus'' of AGN at bright fluxes but widens 
at fainter fluxes, shifting to faint optical magnitudes by $\Delta$(X/O)$\sim$0.5. 
In the hard band, the newly drawn locus behaves similarly to the one in the soft band with the same shift of $\Delta$(X/O)$\sim$0.5 to faint optical magnitudes. 
This shift is consistent with the X/O relation being originally calibrated on soft X-ray selected sources, bright in the optical and also in the X-rays. 
The width of the locus in both bands is constant over 1.5 dex. 
The population of sources outside the newly drawn locus is constant in number in the soft band while becomes larger (in number) at fainter fluxes in the hard band.  

We also performed a linear regression fitting\footnote{The ROBUST$\_$LINEFIT IDL routine is used.} to the logarithm of the X-ray flux 
and the total (aperture corrected) {\it i}-band magnitude. 
In order to have a better statistics at bright fluxes, we also included the XMM-COSMOS sources (outside the \chandra\ area) in the fit using 
the catalog of identifications by Brusa et al. (2010). 
The advantage of including the XMM-COSMOS sources is that the optical photometric catalog is the same, so problems of calibration using different filters 
are not an issue. Moreover, also the good calibration between the 
\chandra\ and \xmm\ fluxes (5-10\%; Paper IV: Lanzuisi et al. in prep.) should not affect the result. Adding the XMM-COSMOS sources creates a sample of 1921 AGN in the soft band and 1597 AGN in the hard band. 
In Figure \ref{xo} we show the fit in both bands computed using the combined sample (red solid line). The fit is consistent with the new locus 
drawn using the median optical magnitude in flux bins. The Spearman rank correlation coefficient of the fit to the combined sample is 
$\rho$=-0.52 for the soft band and $\rho$=-0.48 for the hard band, with a significance level $>$99.9\%. The linear relations found for the 
two bands are {\it i$_{AB}$=}(-5.88$\pm$1.02)+(-1.9$\pm$0.1)$\times$log F$_X$(soft) and 
{\it i$_{AB}$=}(-9.68$\pm$1.39)+(-2.2$\pm$0.1)$\times$ log F$_X$(hard).

\begin{figure*}
\centering
\includegraphics[width=0.49\textwidth]{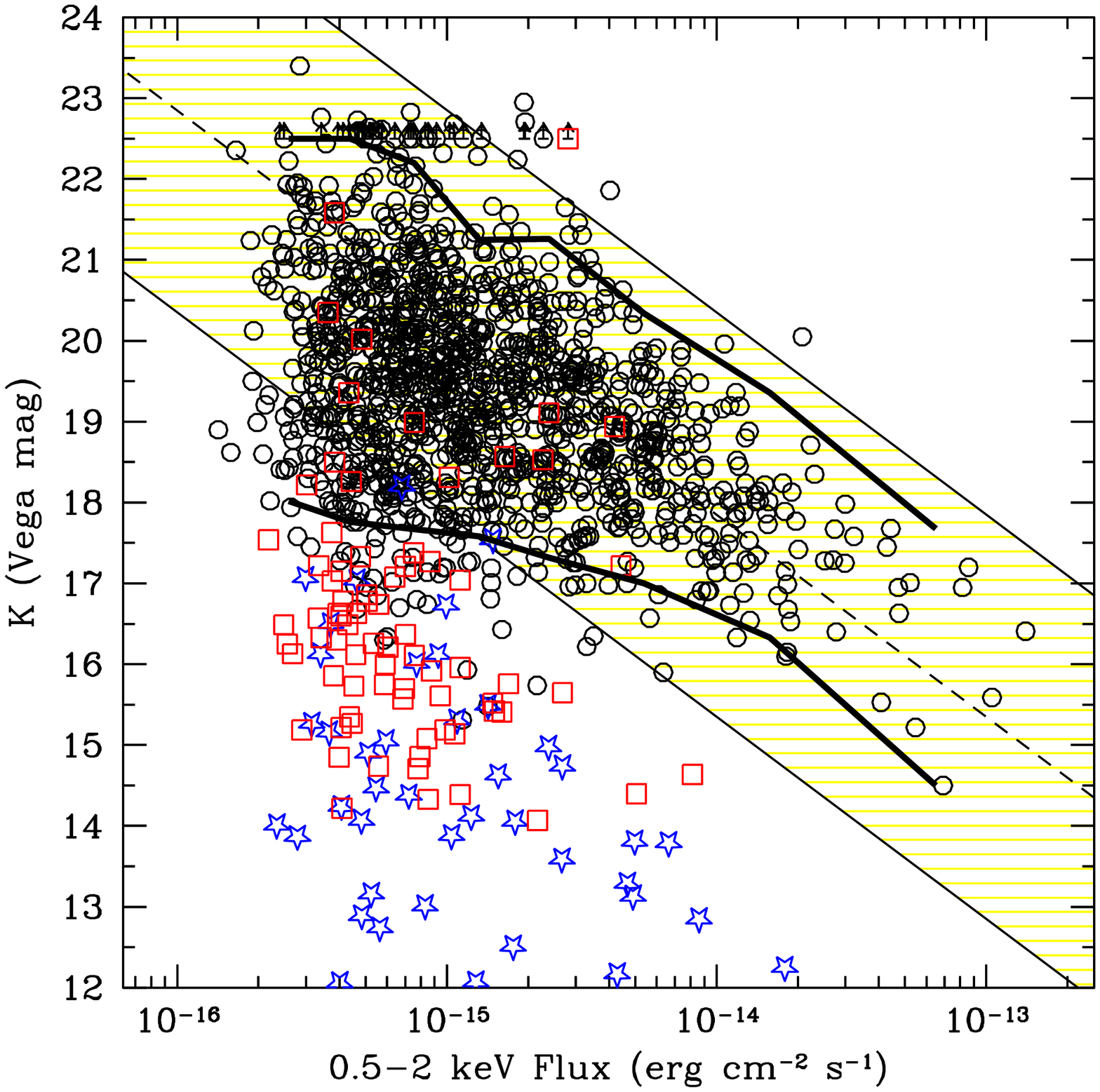}
\includegraphics[width=0.49\textwidth]{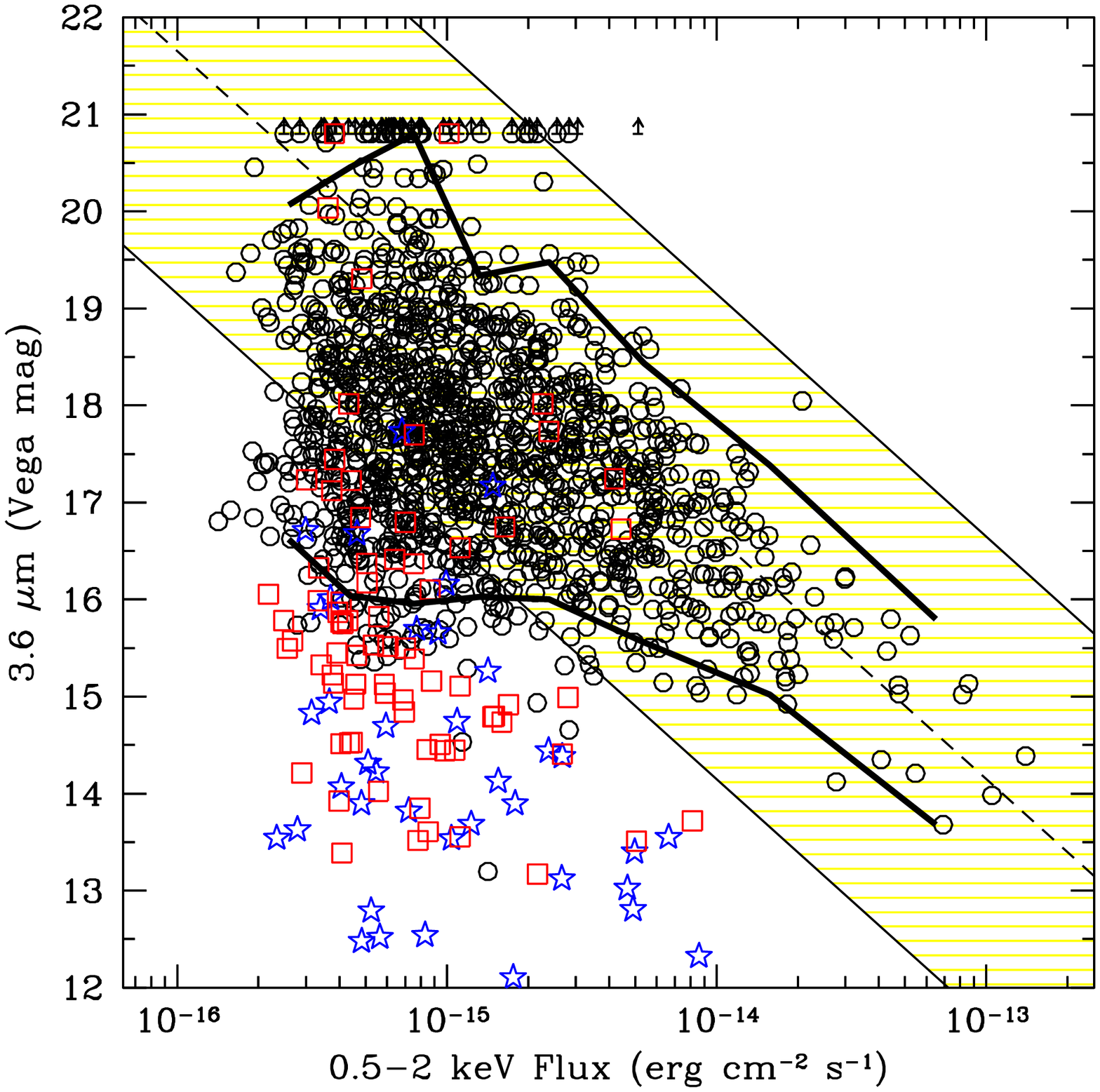}

\caption{\small The X-ray soft flux versus the K (left) and 3.6 $\mu$m (right) total (i.e. aperture corrected) magnitudes for all the X-ray sources. 
The yellow shaded region represents the ``classic locus'' of AGN along the 
correlation X/O=0$\pm$1. Black symbols: C-COSMOS sources. Star symbols represent spectroscopic or 
photometric identified stars. Red and magenta squares represent sources with L$_F<$10$^{42}$ \lum\ in 
C-COSMOS and XMM-COSMOS, respectively. Sources not identified in these bands are represented with a upwards arrows. 
The thick solid curves represent the region including 90\% 
of the AGN population. }
\label{xir}
\end{figure*}

 In Figure \ref{xir} we show the X-ray soft flux versus the total K (left) and 3.6 $\mu$m (right) magnitudes for all the X-ray sources as in Figure \ref{xo}. 
The ``classic locus'', obtained using Eq.~3 and the constant C=6.86 and 7.34 for the K and 3.6 $\mu$m, respectively, is plotted in the figure, as well as the region 
which includes 90\% of the AGN population, as computed for the {\it i} band. 
Similar plots have been shown before in the K band by Mainieri et al. (2002) and 
Brusa et al. (2005) and at 3.6 $\mu m$ by Xue et al. (2011) and Laird et al. (2008). 
In both bands, the spread of the AGN population is reduced by 1.2-2 mags with respect to the {\it i}-band. This suggests a stronger 
coupling between the X-ray flux and the 
near-infrared wavelengths than in {\it i}-band. The higher identification rates in these bands had already hinted at this connection. 
The reduced spread in the near-infrared bands is mainly due to the reduced nuclear extinction which, on the contrary, strongly affects the bluer bands.

\subsubsection{X-ray to optical flux ratio and photometric classification}

\begin{figure}
\centering
\includegraphics[width=0.49\textwidth]{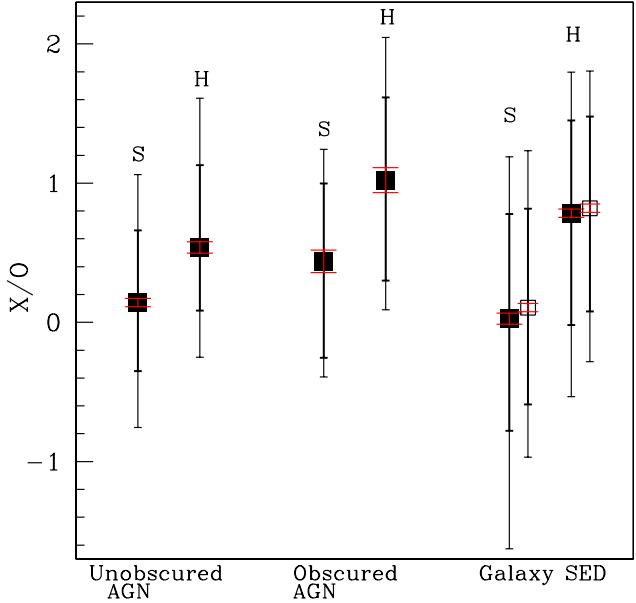}

\caption{\small The median X-ray to optical flux ratio in the soft (S) and hard (H) bands for the sources divided in SED classes (unobscured AGN, obscured 
AGN and galaxy SED sources). 
The open squares represent galaxy SED sources with L$_F>$10$^{42}$ \lum. 
The black bars represent the 68\% and 90\% dispersion. The red error bars represent the median deviation.  
A table with median values and 68\% and 90\% dispersion is available in Table \ref{xo_table}.}
\label{whisker}
\end{figure}

 We also used the photometric classifications, available for a large fraction of the sources via SED fitting, 
to compute the median X/O dividing the sources into three classes (unobscured AGN, obscured AGN and galaxy SED sources). 
This allows us to investigate how the X/O distributions change with X-ray band and source classification. 
The median values of X/O in the hard and soft band for each class along with the 68\% and 90\% ranges are reported in Fig. \ref{whisker} and 
in Table \ref{xo_table}. 
The unobscured AGN fits the ``classic locus'' (-1$<$X/O$<$1 in the soft) and the ``hard band locus'' (-0.53$<$X/O$<$1.47) 
while the obscured AGN have a higher X/O than unobscured AGN and the difference is larger in the hard ($\sim$0.5 dex) than 
in the soft ($\sim$0.3 dex). Their difference in X/O distributions is confirmed also by a Kolmogorov-Smirnov (K-S; P=0.0006). The 90\% range of obscured AGN spreads to X/O$>$2. 

The X/O distribution of galaxy SED sources spans a very large range. In the soft band, the spread at low X/O values is due to the presence of 
low-luminosity normal galaxies (red squares in Fig. \ref{xo}), and it decreases when only luminous sources are considered (open squares in Fig. \ref{whisker}). 
In the hard band, the galaxy SED sources have a median X/O closer to the one of obscured AGN 
and with a smaller spread to low X/O than in the soft band because the contribution of normal galaxies to the hard band X/O 
is negligible (as shown in Fig. \ref{xo}). 
A K-S test confirms that obscured AGN and galaxy SED sources are consistent with being drawn from the same 
parent population (P=0.46 and P=0.58 when only bright galaxy SED sources are included).

 \begin{table*}
\scriptsize
\centering
\caption{\small X/O median values and 68\% and 90\% ranges in the soft and hard band for each SED type. }
\begin{tabular}{l c ccccc}
\hline \hline
Type & Number of sources & \multicolumn{5}{c}{0.5-2 keV}  \\
 & & median X/O & +34\% & -34\% & +45\% & -45\%\\
\hline

Unobscured &   412  & 0.14     &  0.52         &  0.49    &   0.92    &   0.9 \\
Obscured &     87   &0.44      & 0.56	       & 0.69	  &  0.80     &  0.83\\
Galaxy SED  &  864  &  0.03   &       0.75   &    0.81  &      1.16  &   1.65\\
Galaxy SED L$_F>$10$^{42}$\lum  &748 & 0.11 &      0.71&      0.69&       1.13 &       1.07\\
\hline \hline

Type & Number of sources & \multicolumn{5}{c}{2-10 keV}  \\
 & & median X/O & +34\% & -34\% & +45\% & -45\%\\
\hline

Unobscured &   342 & 0.54	&  0.59       &   0.45    &  1.07      &  0.70\\
Obscured &     79  & 1.02	&  0.59       &   0.72    & 1.02	 &  0.93\\
Galaxy SED  &  718 &  0.78    &   0.66        &   0.80   &     1.01   &  1.32\\
Galaxy SED L$_F>$10$^{42}$\lum  &677 & 0.82 &      0.66 &      0.74&      0.99&       1.10\\

\hline \hline
\end{tabular}
\label{xo_table}
\end{table*}

\begin{figure}
\centering
\includegraphics[width=0.5\textwidth]{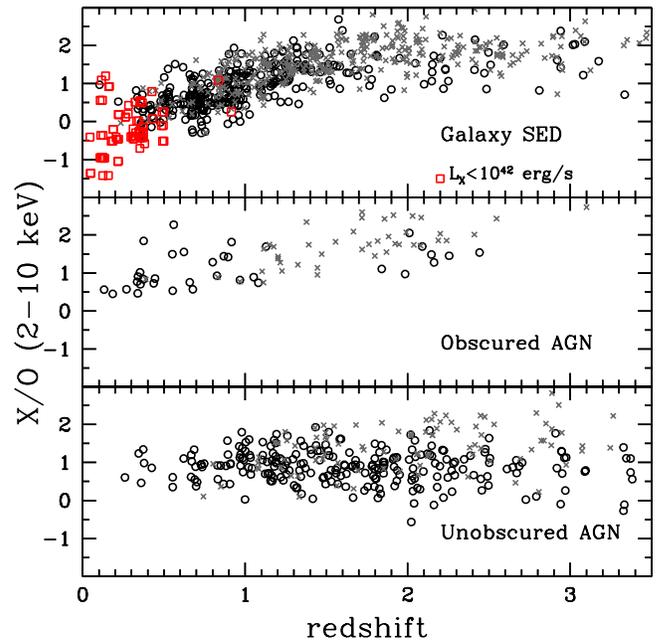}
\caption{\small The X-ray to optical flux ratio in the hard band versus the redshift, dividing the sources on the basis of the optical best fitting template. 
Black circles represent sources with a spectroscopic identification, gray crosses those without. The red squares represent sources with L$_F<$10$^{42}$ \lum.  }
\label{xozeta}
\end{figure}

As it has been already argued and observed by several authors, obscured accretion at high redshift and the fact that the optical flux 
is more affected by absorption than the X-ray flux are responsible for high X-ray to optical flux ratios.
When redshifting the SED of an obscured AGN to progressively higher redshifts the k-corrections in the optical and X-ray band work
in opposite directions. 
The X-ray flux, with a typical extremely hard spectral slope, has a positive k-correction, increasing with redshift and boosting the X-ray flux. 
On the contrary, in the optical, their SED is dominated either by galaxy starlight or by dust 
attenuated nuclear emission or by a combination of both. For these reasons, the k-correction works in the opposite direction 
with respect to the X-ray one: moving to high redshifts, the faint rest frame optical/UV emission is shifted to the {\it i} 
band explaining the progressively faint optical magnitudes (see a similar discussion in Brusa et al. 2010 and Comastri et al. 2003). 
The result is that the change in observed X-ray flux with increasing redshift is smaller than the change in optical {\it i}-band magnitude 
and thus obscured AGN and galaxy SED sources at high redshift have higher X/O.
Figure \ref{xozeta} shows that galaxy SED sources occupy the same region of the X/O versus redshift plane of obscured AGN and extend to higher-redshifts and 
larger X/O. This suggests that galaxy SED sources are mainly obscured AGN at high redshift. 

Unobscured sources have a similarly steep spectral slopes in both the 
optical and X-rays ($\alpha \sim$1), thus the two k-corrections are similar. For this reason, the spread in X/O in the $\pm$1 range 
(Fig. \ref{whisker}) remains constant with redshift (Fig. \ref{xozeta}) and it is merely due to the spread in luminosity. 
Consequently, there is no correlation between X/O and luminosity for unobscured AGN, while it has been widely shown that X-ray to optical flux ratio 
correlates with hard X-ray luminosity for obscured AGN (Fiore et al. 2003, Eckart et al. 2006, Brusa et al. 2010).

\begin{figure}
\centering
\includegraphics[width=0.5\textwidth]{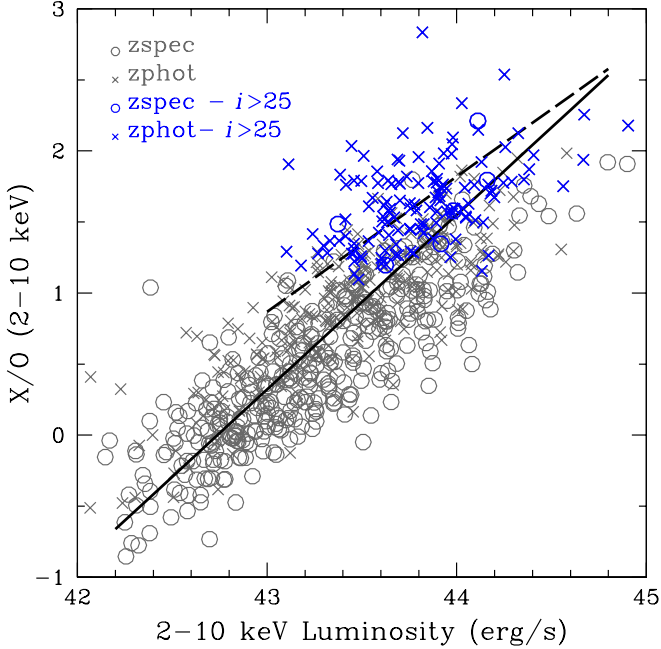}
\caption{\small The X-ray to optical flux ratio in the hard band versus the hard band X-ray luminosity. Circles represent sources with a spectroscopic 
identification, crosses those without. The blue sources have {\it i}$>$25. The solid line is the fit to the total sample. 
The dashed line is the fit to the faint optical sources.  }
\label{lxxo}
\end{figure}

In Figure \ref{lxxo}, the hard X-ray luminosity and the X/O (computed in the hard band; Fiore et al. 2003) is plotted for 
the obscured AGN and the galaxy SED sources at luminosity $>$10$^{42}$ \lum. The sample consists of 762 sources. The correlation (with a slope of 1.2; black solid line) 
is quite strong over 3 orders of magnitude in luminosity (Spearman rank correlation coefficient $\rho$=0.8) and it is stable even when the sample is divided  
between sources with spectroscopic and photometric redshifts. 
This correlation again supports the fact that there is a good coupling between the galaxy and the central 
BH over a wide range of X-ray luminosities. In the past, it has been pointed out that at faint optical magnitudes, the two quantities do not 
correlate (Barger et al. 2005, Civano et al. 2005), or follow a different relation. Brusa et al. (2010) found the same behavior for the faint (R$>$23) 
XMM-COSMOS sources. We considered here the sub-samples of sources at {\it i}$>23$ (450 sources) and  {\it i}$>25$ (blue symbols in Fig. \ref{lxxo}). 
We find that the first group follows approximately the same correlation of the total sample, while, at fainter magnitudes, 
there is still a correlation but with different normalization and slope (0.9; dashed line).  However, a larger sample of faint sources is necessary to verify 
if the correlation holds, suggesting a different population of sources, or it is mainly a spread towards high X/O due to the detection of low-luminosity host galaxies 
at higher redshifts. 
Thus this correlation can still be used to estimate redshifts for optically bright sources, as 
suggested in the past (Fiore et al. 2003), but some caveats have to be considered for faint sources.

\subsubsection{X-ray to optical flux ratio and morphology}

Recently, Povi{\'c} et al. (2009a, b) found an anti-correlation between the X/O ratio and the concentration parameter (C; Abraham et al. 1994), which represents 
how the source light is distributed among the galaxy pixels. X/O ratio could be anti-correlated with the production efficiency of the BH (i.e. the Eddington ratio; Kelly et al. 2008, 
Povi{\'c} et al. 2009a, Trump et al. 2011). Thus the anti-correlation could be due to a dependence of the Eddington ratio and the galaxy mass, which strongly 
correlates with C (Graham et al. 2001a,b).   

\begin{figure}
\centering
\includegraphics[width=0.5\textwidth]{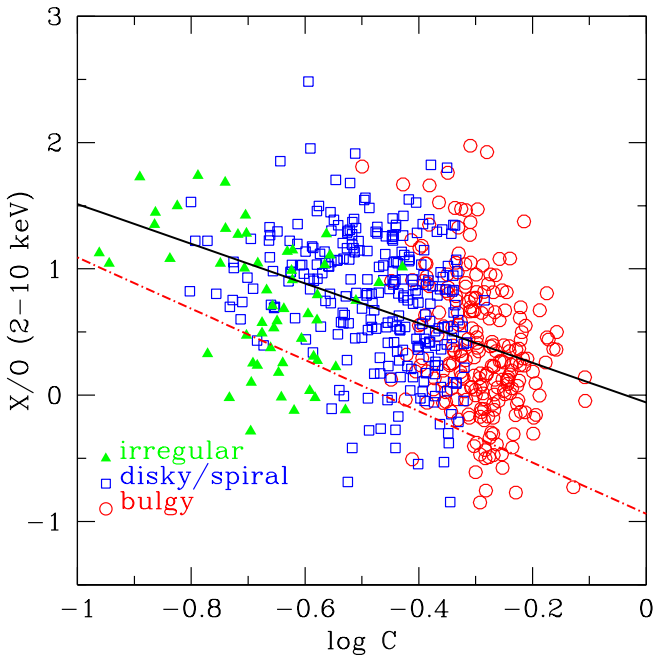}
\caption{\small The X-ray to optical flux ratio in the hard band versus the concentration parameter measured in the HST ACS images. Red 
circles represent sources with a {\it bulgy} morphology, blue square sources with a {\it disky}/{\it spiral} morphology, green triangles sources with 
an {\it irregular} morphology.
The black solid line is the linear fit. The red dot-dashed line is the Povi{\'c} et al. (2009b) fit.  }
\label{rapcon}
\end{figure}

We correlated the C-COSMOS counterparts catalog with the Tasca et al. (2009) catalog which reports the 
morphological parametric quantities, computed using Morpheus 2005 (Abraham et al. 2007) on HST-ACS images, for several thousands COSMOS sources to {\it i}=22.5. 
In order not to be contaminated by unobscured AGN, for which the presence of a bright nuclear source can 
affect the determination of the concentration index (Gabor et al. 2009), we excluded 
from the sample sources classified as point-like from the ACS catalog (Leauthaud et al. 2007) and also sources classified as unobscured from their SED. Also, 
to avoid the contamination by normal star-forming galaxies we excluded from the sample sources with luminosity in the full band $<$10$^{42}$ \lum. 
In Figure \ref{rapcon}, the X/O computed in the hard band and $log$(C) are plotted for the 561 sources with the above selection.

We find a significant anti-correlation (Spearman coefficient $\rho$=0.45; solid line in Figure \ref{rapcon}) between 
the two quantities. The scatter around the correlation is large, however the significance is the same of Povi{\'c} et al. (2009). 
The best fit relation is $X/O=-0.057-1.57\times C$. 
We note that Povi{\'c} et al. (2009) uses different X-ray band and optical magnitude to compute the X/O ratio, so 
the normalization of their fit is quite different (dot-dashed line in Figure \ref{rapcon}) and also their slope is 
steeper but this could be due to the limited number of sources in the 
Povi{\'c} et al. (2009b) sample ($\sim$100 sources). The large range of X/O covered implies that this anti-correlation 
is valid for a sample spanning a large range of redshifts 
and luminosities, thus being representative of the entire sample. 

We used the same morphological catalog to classify the sources in 3 different classes: bulgy (red circles), disky/spiral (blue squares) and irregular (cyan triangles). 
We find that most of the sources (50\%) are classified as disky/spiral, 40\% as bulgy galaxies and only 10\% of the sources 
show an irregular morphology. 
This result is in agreement with recent findings (e.g., Gabor et al. 2009, Cisternas et al. 2010, Schawinski et al. 2011, Kocevski et al. 2011, Kriek et al. in prep.), 
suggesting that secular processes rather then major mergers govern a significant fraction of the obscured BH growth in the luminosity regime sampled by X-ray surveys.

We note that the median X/O value for the 3 classes increases with decreasing concentration and that the fraction of sources 
at L$_X>$10$^{43.5}$ \lum is larger in irregular sources (45\%) than in disky/spiral (36\%) and bulgy (25\%) galaxies. 
Thus, sources with irregular morphologies tend to have higher X-ray luminosities ($>10^{43.5}$ \cgs in the hard band) and 
higher X/O (indicating higher obscuration) than sources with undisturbed (spiral or bulgy) morphologies. 
This observed behavior is currently favored by models of merger-driven AGN fueling (e.g., Hopkins et al. 2008), 
where the obscured accretion phase should happen in sources with an irregular 
morphology due to major mergers, suggesting that these models are valid in the high luminosity regime only.

\subsection{Optical to X-ray Color Diagram}
To further study the nature of the sources dominated by the host galaxy light in the optical (galaxy SED sources and obscured AGN), 
the X-ray properties were analyzed and compared with the optical properties.
 
The hardness ratio (HR) provides a first, approximate indication 
of the shape of the X-ray spectra. 
HR is defined as HR = (H -S)/(H + S), where H is the number of counts in the hard band and S is the number of counts in the soft band.
Comparing the column densities derived from the spectral analysis of the brightest C-COSMOS sources (Paper IV) with their HRs, we estimated that HR=-0.2 is 
an appropriate value to roughly separate sources with significant obscuration (N$_H >10^{22}$ cm$^{-2}$) from effectively unobscured sources.
This value is consistent with what has been usually used in the literature in other \chandra\ surveys (e.g., Wilkes et al. 2006), even if we 
have to keep in mind that it is only an approximation, given that the relation between X-ray column density and HR is strongly redshift dependent 
(see Civano et al. 2011, Figure 1) and the statistics for faint sources are poor .
For the sources not detected in the hard or soft band we computed upper or lower limits on the HR,  
using for each band the smallest number of counts detected in the field (3.7 in the soft band and 4.5 in the hard band). 

\begin{figure}
\centering
\includegraphics[width=0.49\textwidth]{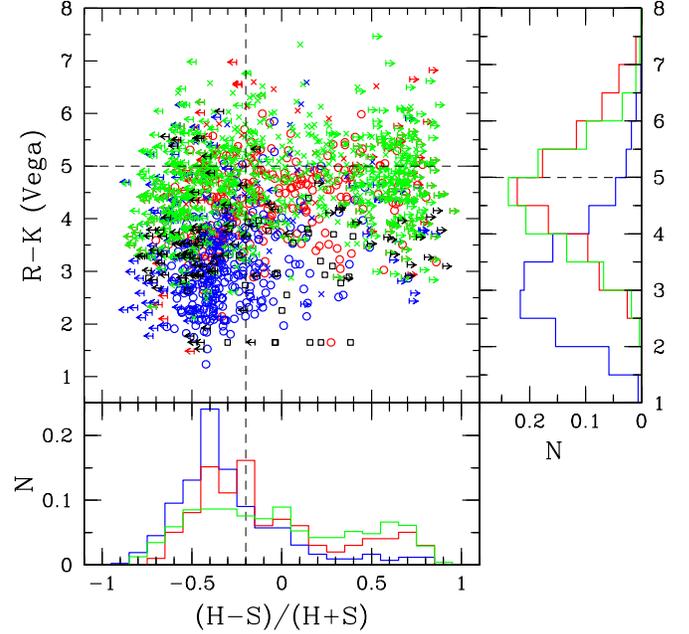}
\caption{\small 
The R-K color in the Vega system versus the X-ray hardness ratio. 
Circles represent sources with a spectroscopic identification, crosses those without. Arrow symbols represent sources without a detection 
in the soft or hard band for which a limit has been estimated. Blue=unobscured AGN, 
red=obscured AGN, green= galaxy SED sources. The black squares and arrows represent sources with $L_F<$10$^{42}$ \lum. 
In the right and bottom histograms, the distributions of the two quantities in photometric classes are plotted. 
The HR distributions take into account also upper limits. }
\label{rkhr}
\end{figure}

In the optical and infrared bands, the $R-K$ color has been used in the past to select obscured sources in X-ray surveys (Brusa et al. 2005) and 
it has been shown that a correlation exists between X/O and $R-K$ (Brusa et al. 2010, Rovilos et al. 2010). 
In Figure \ref{rkhr} (left panel), the R-K color (computed using aperture photometry) is plotted versus the HR for all the sources, 
together with the distributions of the two quantities, dividing the sources in the three photometric types. 
The sources with X-ray luminosity $<$10$^{42}$ \lum\ (red squares) are scattered across the plane without showing a correlation between 
HR and optical colors. These sources have been excluded from the histograms in Figure \ref{rkhr} (left panel). 

Unobscured AGN (blue histogram) have on average blue color ($\langle R-K \rangle$ =3.27) in the optical and soft spectra in the X-ray 
($\langle HR \rangle$ =-0.29, including upper limits), even if a tail at higher HR and $R-K$ is present, 
in agreement with other findings of red quasars with broad lines (Glikman et al. 2007, Urrutia et al. 2008, Young et al. 2009).  
The sources fitted with a galaxy template have HR and $R-K$ distributions 
consistent with those of obscured AGN, with $\langle HR \rangle$=-0.01 and -0.08, and $\langle R-K \rangle$ =4.65 and 4.83, respectively. Similar numbers are obtained 
when upper limits are not considered. 
If BLAGN and non-BLAGN are considered, the HR and $R-K$ colors are consistent with those found if the photometric classification is used 
($\langle HR \rangle$ =-0.31 and -0.04, and $\langle HR \rangle$  =3.1 and 4.3).

\section{Summary}
\label{discussion}

In this paper we have presented the identification procedure, the catalog and some basic properties of the X-ray point-like sources detected in the C-COSMOS project. 
We summarize the most important results in the following: 
 
\begin{itemize}

\item The procedure for the association of the \cps, performed via the likelihood ratio technique in 3 different bands (i, K, 3.6$\mu$m), 
has produced a very high identification rates (97\%), 
after matching the results in the 3 bands and also thanks to the good correlation of near-infrared data with X-ray data at the depth of C-COSMOS.

\item Optical and near-infrared information is available for all the \cps\ in the sample but 10 objects, two of which are truly empty fields.  

\item Secure spectroscopic redshifts with two or more emission and/or absorption lines are available for 
906 sources. 35\% of them are identified as BLAGN and 59\% do not show broad 
lines but only narrow lines or absorption lines. The remaining sources are spectroscopically identified stars. 

\item Photometric redshifts are available for 1693 
sources. According to the photometric classification, based on the fitting of the SED, most of the sources are best fitted by a pure galaxy template (65\%), 
though in the whole sample only $\sim$10\% of the sources have luminosity $<$10$^{42}$ \lum\, possibly due to non-nuclear emission.
 
\item  C-COSMOS occupies a sweet spot in the X-ray survey ``wedding cake'': the perfect match between the X-ray and optical/near-infrared depth allows us 
to classify the z=0-5 X-ray sources just using their photometric data; the large area provides sizable samples of rare sources 
(e.g., high--z sources) and allows us to recognize a very large sample of obscured sources (71\% on the basis of the photometric SED fitting and the X-ray luminosity), 
either best fitted in the optical by an obscured (hybrid) AGN template or by a galaxy template. These obscured sources start to dominate the whole 
X-ray population below fluxes of 4$\times$10$^{-15}$ \cgs\ (0.5-2 keV). The optical and X-ray properties (redshifts, X/O, HR, $R-K$ and N$_H$) 
confirm that most of them are highly-obscured z=1--2 AGN. The X-ray band remains the only band able to isolate this population of obscured sources which 
in the optical show a normal galaxy SED. 

\item Thanks to the large sample of AGN in the C-COSMOS survey, we revised, both in the soft and in the hard band, the X-ray to optical flux ratio 
locus, originally defined in the soft band. We combined the C-COSMOS and XMM-COSMOS catalogs, providing a sample of 
2214 sources in the soft band and 1676 in the hard band, spanning 2 order of magnitudes in X-ray flux. 90\% of the sources lie in a region 
slightly offset from the ``classic locus'' in the soft band, while in the hard band the new locus is offset by $\Delta$(X/O)$\sim$0.5 
with respect to the ``hard band locus''. 
We also provide the best fits to the X-ray flux and total {\it i}-band magnitude for the AGN sample in both bands.

\item We also presented the X-ray to K and 3.6$\mu$m flux ratios which shows a smaller spread of the sources, indicating a strong correlation of X-rays with near-infrared 
bands for both obscured and unobscured AGN, as already pointed in the identification procedure.

\item The X/O ratio correlates with several optical, infrared and X-ray properties of X-ray sources. We confirm the presence of a correlation in the 
hard band between the X/O ratio and the X-ray luminosity for obscured and galaxy SED sources, as found by several authors, confirming the tight correlation 
between the galaxy and the BH properties over a wide X-ray luminosity range and to a magnitude limit of {\it i}$=25$. This correlation has been 
used in the past to estimate the X-ray luminosity and thus the redshift of X-ray sources, just using the observed fluxes. While the fit to the bright sample 
has a slope consistent with previous work, the fit to the faint sample ({\it i}$>25$) has a flatter slope and different normalization, thus putting a limit on 
the validity of this relation as good estimator of redshift and luminosity at faint magnitudes.

\item An anti-correlation between the X/O in the hard band and the concentration parameter has been found, confirming the results of Povi{\'c} et al. (2009), 
by using a 6 times larger sample. We also find that most of the high X/O sources, thus obscured, are classified as disky/spiral galaxies suggesting that 
secular processes govern a significant fraction of the obscured BH growth, on a wide range of X-ray luminosities (10$^{43}$-10$^{44.5}$\lum) and redshift. 
The only sources with an irregular morphology are those at the brightest luminosities, in agreement with the predictions of merger-driven fueling scenarios.
The next step will be to compute galaxy masses, by using a careful SED fitting which takes into account the nuclear component in the optical band as in Lusso et al. (2011), 
and confirm the correlation between X/O and the galaxy mass. 

\end{itemize}

We began the Chandra COSMOS project in the belief that it represented
a sweet spot in the inevitable depth versus area trade-off, and the work presented here vindicates this approach. 
Other C-COSMOS results (e.g. Capak et al. 2011) have shown that this survey is also powerful at identifying large scale structures at z$>$4 spanning 
a large area of the sky ($>$15$^{\prime\prime}$). Given that there is twice as much area in the full COSMOS field with
identical, and increasing, multi-wavelength coverage, we feel that the best path for deep/wide extragalactic \chandra\ future surveys would be to
complete the COSMOS field (2 deg$^2$) to at least the present depth.\\

\begin{acknowledgements}
The authors thanks the referee for the useful suggestions which helped to improve the paper. FC thanks R. D'Abrusco, A. Goulding, M. Mignoli and M. Povi{\'c} 
for inspiring discussions and useful comments. This work was supported in part by NASA Chandra grant
number GO7-8136A (ME, FC, HH), the Blancheflor Boncompagni 
Ludovisi foundation (FC) and the Smithsonian Scholarly Studies (FC). 
In Italy this work is supported by ASI/INAF contracts I/009/10/0 and
I/088/06 and by PRIN INAF 2010 ``From the dawn of galaxy formation to the peak of the mass assembly''.
MS acknowledge support by the German Deutsche Forschungsgemeinschaft, DFG Leib- niz Prize (FKZ HA 1850/28-1).
In Mexico, this work has been supported by CONACyT Grant 83564 and UNAM/DGAPA  PAPIIT grant IN110209. 
This work is based on observations made with the \chandra\ X-ray satellite, 
with ESO Telescopes at the La Silla or Paranal Observatories under programme ID 175.A-0839.

\end{acknowledgements}

\end{document}